\documentstyle[prd,aps,epsf,amsfonts,amssymb,amsmath]{revtex}

\def\overlinen#1{{\bar{#1}}}
\def\calZ{{\mathcal{Z}}}
\def\calZ{{{Z}}}
\def\calN{{\mathcal{N}}}
\def\calS{{\mathcal{A}}}
\def\mB{\mathbb{B}}
\def\|{|\!|}
\def\mB{{B}}
\newcommand{\be}{\begin{equation}}\newcommand{\ee}{\end{equation}}
\newcommand{\bea}{\begin{eqnarray}}\newcommand{\eea}{\end{eqnarray}}
\newcommand{\brr}{\begin{array}}\newcommand{\err}{\end{array}}
\newcommand{\bit}{\begin{itemize}}\newcommand{\eit}{\end{itemize}}
\newcommand{\ben}{\begin{enumerate}}\newcommand{\een}{\end{enumerate}}

\newcommand{\ide}{1\hspace{-1mm}{\rm I}}

\def\noi{\noindent}

\def\1{{_{1}}}\def\2{{_{2}}}
\def\boldsymbol#1{{\bf #1}}%
\begin{document}

\title{Path Integral Approach to 't\,Hooft's Derivation of
Quantum from Classical Physics
}

\vspace{5mm}

\author{Massimo Blasone${}^{\dag}$,  Petr Jizba${}^{\ddag}{}^{\flat}$, and
Hagen Kleinert${}^{\natural}$}

\address{ $ $\\[2mm]
${}^{\dag}$
Dipartimento di  Fisica, Universit\`a di Salerno,
Via S.Allende, 84081 Baronissi (SA) - Italy
\\
${}^{\ddag}$ Institute for Theoretical Physics, University of
Tsukuba, Ibaraki 305-8571, Japan\\
${}^{\flat}$ FNSPE, Czech Technical University, Brehova 7, 115 19
Praha 1, Czech Republic\\
${}^{\natural}$ Institut f\"{u}r Theoretische Physik, Freie
Universit\"{a}t Berlin, Arnimallee 14 D-14195 Berlin, Germany
\\ [2mm] E-mails: blasone@sa.infn.it, p.jizba@fjfi.cvut.cz,
kleinert@physik.fu-berlin.de
 }

\maketitle

\vspace{1mm}
\begin{center}
{\small \bf Abstract}
\end{center}
\vspace{-7mm}
\begin{abstract}
\hspace{-10.5mm}

We present a path-integral formulation of 't~Hooft's derivation of
quantum from classical physics. The crucial ingredient of this
formulation is Gozzi {\em et al.} supersymmetric path integral of
classical mechanics. We quantize explicitly two simple classical
systems: the planar mathematical pendulum and the R\"{o}ssler
dynamical system.

~\\

\vspace{-3mm}
\end{abstract}


\section{Introduction}

In recent decades, various classical, i.e.,
 deterministic  approaches
to quantum theory have been proposed. Examples are Bohmian
mechanics~\cite{Bohm1}, and the stochastic quantization procedures
of Nelson~\cite{Nelson1}, Guerra and Ruggiero~\cite{Guerra1}, and
Parisi and Wu~\cite{Parisi1,Huffel}. Such approaches are finding
increasing interest
 in the physics
community. This might be partially ascribed to the fact that such
alternative formulations help in explaining some quantum phenomena
that cannot be easily explained
 with the usual formalisms. Examples
are multiple tunneling~\cite{Jona-Lasinio}, critical phenomena at
zero temperature~\cite{Ruggiero1}, mesoscopic physics and quantum
Brownian oscillators~\cite{Rugierro2}, and quantum-field-theoretical
regularization procedures which manifestly preserve
 all symmetries of the
bare theory such as gauge symmetry, chiral symmetry, and
supersymmetry~\cite{reg}. They allow one to quantize gauge fields,
both Abelian and non-Abelian, without gauge fixing and the ensuing
cumbersome Faddeev-Popov ghosts~\cite{FPG}, etc..

\vspace{3mm}


The primary objective of a reformulation of quantum theory in the
language of classical, i.e.,  deterministic theory is basically
twofold. On the formal side, it is hoped that this will help in
attacking quantum-mechanical problems from a different direction
using hopefully more efficient mathematical techniques than the
conventional ones. Such techniques may be based on stochastic
calculus, supersymmetry, or various new numerical approaches (see,
e.g., Refs.~\cite{Huffel,Pain} and citations therein). On the
conceptual side, deterministic scenarios are hoped to shed new light
on some old problems of quantum mechanics, such as the origin of the
superposition rule for amplitudes and  the theory of quantum
measurement. It may lead to new ways of quantizing chaotic dynamical
systems, and ultimately a long-awaited  consistent theory of quantum
gravity. There is, however, a price to be paid for this; such
theories must have a built-in nonlocality
 to
escape problems with Bell's inequalities. Nonlocality may be
incorporated in numerous ways --- the Bohm-Hiley quantum
potential~\cite{Bohm1,Bohm2}, Nelson's osmotic
potential~\cite{Nelson1}, or Parisi and Wu's {\em fifth--time\/}
parameter~\cite{Parisi1,Huffel}.

\vspace{3mm}

Another deterministic access to quantum-mechanical systems was
recently proposed by 't Hooft ~\cite{tHooft,tHooft3} with subsequent
applications in
Refs.\cite{BJV1,tHooft22,Halliwell:2000mv,cvb,BMM1,Banajee,Elze2}.
It is motivated by black-hole thermodynamics (and particularly by
the so-called {\em holographic principle\/}~\cite{tHooft2,Bousso}),
and hinges on the concept of {\em information loss\/}. This and
certain accompanying non-trivial geometric phases are able to
explain the observed non-locality in quantum mechanics. The original
formulation has appeared in two versions: one involving a  discrete
time axis~\cite{tHooft22}, the second continuous
times~\cite{tHooft3}. The goal of this paper is to discuss further
and gain more understanding of the latter model. The reader
interested in the discrete-time model may find some practical
applications in Refs.~\cite{BJV3,Elze}. It is not our purpose to
dwell on the conceptual foundations of 't\,Hooft's proposal. Our aim
is to set up a possible useful alternative formulation of
't\,Hooft's model and quantization scheme that is based on path
integrals \cite{Pain}. It makes use of Gozzi {\em et al.}
path-integral formulation of classical
mechanics~\cite{GozziI,GozziII} which appears to be a natural
mathematical framework for such a discussion. The condition of the
information loss, which is basically a first-class subsidiary
constraint, can then be incorporated into path integrals  by
standard techniques. Although 't\,Hooft's procedure differs in its
basic rationale from stochastic quantization approaches, we show
that they share a common key feature, which is
 a hidden BRST invariance, related to
 the so-called Nicolai map~\cite{Nikoloai1}. To be specific, we shall
apply our formulation to two classical systems: a planar
mathematical pendulum and the simplest deterministic chaotic system
--- the R\"{o}ssler attractor. Suitable choices of the ``loss of
information" condition then allow us to identify the emergent
quantum systems with a free particle, a quantum harmonic oscillator,
and a free particle weakly coupled to Duffing's oscillator.

\vspace{3mm}

Our paper is organized as follows. In Section \ref{SEc2} we quantize
't\,Hooft's Hamiltonian system by expressing it in terms of a path
integral which is singular due to the presence of second-class
primary constraints. The singularity is removed with the help of the
Faddeev-Senjanovic prescription~\cite{Fad,Senj}. It is then shown
that the fluctuating system produces a classical partition function.
In Section \ref{SEc3} we briefly review Gozzi {\em et al.}
path-integral formulation of classical mechanics in configuration
space. The corresponding phase-space formulation is more involved
and will not be considered here. By imposing the condition of a
vanishing ghost sector, which is characteristic for the underlying
deterministic system, we find that the most general Hamiltonian
system compatible with such a condition is the one proposed by
't\,Hooft. In Section \ref{SEc4} we introduce 't\,Hooft's constraint
which expresses the property of information loss. This condition not
only explicitly breaks the BRST symmetry but, when coupled with the
Dirac-Bergmann algorithm, it also allows us to recast the classical
generating functional into a form representing a proper
quantum-mechanical partition function. Section V is devoted to
application of our formalism to practical examples. We conclude with
Section VI.
For the reader's convenience the paper is supplemented with four
appendixes which clarify some finer mathematical points needed in
the paper.


\section{Quantization of 't\,Hooft's Model} \label{SEc2}

Consider the class of systems described by Hamiltonians of the form
\begin{eqnarray}
H = \sum_{a=1}^N p_a f_a({\boldsymbol{q}})\, . \label{eq.1.1}
\end{eqnarray}
Such systems emerge in diverse physical situations, for example,
Fermi fields, chiral oscillators~\cite{Banajee}, and noncommutative
magnetohydrodynamics~\cite{Jackiw}. The relevant example in the
 present context is the use of (\ref{eq.1.1})
 by 't\,Hooft to formulate his
{\em deterministic quatization\/} proposal~\cite{tHooft}.

\vspace{3mm}

An immediate problem with the above Hamiltonian is its unboundedness
from below. This is due to the absence of a leading kinetic term
quadratic  in the momenta $p_ a ^2/2M$, and we shall dwell more on
this point in Section~\ref{SEc4}. The equations of motion following
from Eq.(\ref{eq.1.1}) are
\begin{eqnarray}
\dot{q}_a \ = \ f_a({\boldsymbol{q}})\, , \;\;\;\; \dot{p}_a \ = \
- p_a \frac{\partial f_a({\boldsymbol{q}})}{\partial q_a}\, .
\label{eq.1.1.1}
\end{eqnarray}
Note that the equation for $q_a$ is autonomous, i.e., it is
decoupled from the conjugate momenta $p_a$. The absence of a
quadratic term makes it impossible to find a Lagrangian via a
Legendre transformation. This is because the system is singular
--- its Hess matrix $H^{ab }\equiv \partial ^2H/\partial p_
a\partial p _b$ vanishes.

\vspace{3mm}

A Lagrangian yielding the equations of motion (\ref{eq.1.1.1}) can
nevertheless be found, but at the expense of doubling the
configuration space by introducing additional auxiliary variables
$\overlinen q_ a ~(a=1,\dots, N)$. This {\em extended} Lagrangian
has the form
\begin{eqnarray}
\overlinen L \ \equiv \ \sum_{a=1}^N \left[\bar{q}_a \dot{q}_a -
\bar{q}_a
f_a({\boldsymbol{q}})\right] 
\,
 \label{lag1}
\end{eqnarray}
%
and it allows us to define canonically  conjugate momenta in the
usual way: $p_a \equiv  \partial \overlinen L/\partial \dot{q}_a,~
\overlinen{p}_a \equiv
\partial \overlinen L/\partial \dot{\overlinen{q}}_a$.
A Legendre transformation produces the Hamiltonian
\begin{eqnarray}
\overlinen H(p_a, q_a, {\overlinen{p}}_a, {\overlinen{q}}_a) =
\sum_{a=1}^N p_a \dot{q}_a + {\overlinen{p}}_a
\dot{{\overlinen{q}}}_a - L = \sum_{a=1}^N \bar{q}_a
f_a({{\boldsymbol{q}}})\, . \label{2.4}
\end{eqnarray}
The rank of the Hess matrix is zero which gives rise to $2N$ primary
constraints, which can be chosen as:
\begin{eqnarray}
\phi_1^a = p_a - \overlinen{q}_a \ \approx \ 0\, , \;\;\;\;\;\;
\phi_2^a = \overlinen{p}_a \ \approx \ 0\, . \label{2.5}
\end{eqnarray}
The use of the symbol $\approx$ instead of $=$ is due to
Dirac~\cite{Dir} and it has a special meaning: two quantities
related by this symbol  are equal after all constraints have been
enforced. The system has no secondary constraints (see Appendix A).
The matrix formed by the Poisson brackets of the primary
constraints,
\begin{eqnarray}
\{\phi_1^a(t) , \phi_2^b(t) \} \ = \ - \delta_{a b}\, , \;\;\;
\label{2.10}
\end{eqnarray}
has a nonzero determinant, implying that all constraints are of the
second class. Note that on the constraint manifold the {\em
canonical} Hamiltonian (\ref{2.4}) coincides with 't\,Hooft's
Hamiltonian (\ref{eq.1.1}).

\vspace{3mm}

To quantize 't\,Hooft's system we
utilize the general  Faddeev-Senjanovic path integral
formula~\cite{Fad,Senj} for time evolution amplitudes%
\footnote{Other path-integral representations of systems with
second-class constrains such as that of Fradkin and
Fradkina~\cite{fradkin} would lead to  the same result
(\ref{eg.1.2}). }
\begin{eqnarray}
\langle {\boldsymbol{q}}_2,t_2| {\boldsymbol{q}}_1, t_1 \rangle =
{\calN} \int {\mathcal{D}}{\boldsymbol{p}}
{\mathcal{D}}{\boldsymbol{q}} \ \sqrt{\left|\det \|\{\phi_i ,
\phi_j \} \|       \right|} \ \prod_i\delta[\phi_i]\ \exp \left\{
\frac{i}{\hbar } \int_{t_1}^{t_2} dt \left[ {\boldsymbol{p}}
\dot{{\boldsymbol{q }} } - \overlinen H({\boldsymbol{q}},
{\boldsymbol{p}} )\right] \right\}\, . \label{frad}\end{eqnarray}
Using the shorthand notation $\phi_i = \phi_1^1, \phi_2^1,\,
\phi_1^2, \phi_2^2,\, \ldots, \phi_1^N, \phi_2^N ~(i=1,\dots,2N)$,
Eq.(\ref{frad}) implies in our case that
\begin{eqnarray}
\langle {\boldsymbol{q}}_2,t_2| {\boldsymbol{q}}_1, t_1 \rangle
&=& {\calN} \int {\mathcal{D}}{\boldsymbol{p}} {\mathcal{D}}
{\boldsymbol{q}} {\mathcal{D}} \overlinen{{\boldsymbol{p}}}
{\mathcal{D}}\overlinen{{\boldsymbol{q}}} \
\delta[{\boldsymbol{p}} - \overlinen{{\boldsymbol{q}}}] \,
\delta[\overlinen{{\boldsymbol{p}}}]\ \exp\left\{\frac{i}{\hbar }
\int_{t_1}^{t_2} dt\,[{\boldsymbol{p}} \dot{{\boldsymbol{q}}} +
\overlinen{{\boldsymbol{p}}} \dot{\overlinen{{\boldsymbol{q}}}} -
\overlinen H({\boldsymbol{q}}, \overlinen{{\boldsymbol{q}}},
{\boldsymbol{p}}, \overlinen{{\boldsymbol{p}}} )]
\right\}\nonumber
\\
&&~ \nonumber \\
&=& {\calN}  \int_{{\boldsymbol{q}}(t_1) =
{\boldsymbol{q}}_1}^{{\boldsymbol{q}}(t_2) = {\boldsymbol{q}}_2}
{\mathcal{D}}{\boldsymbol{q}}
{\mathcal{D}}\overlinen{{\boldsymbol{q}}} \ \exp\left[\frac{i}{\hbar }
\int_{t_1}^{t_2}\overlinen L({\boldsymbol{q}}, \overlinen{{\boldsymbol{q}}},
\dot{{\boldsymbol{q}}}, \dot{\overlinen{{\boldsymbol{q}}}}) \ dt
\right]\nonumber \\
&&~ \nonumber \\
&=& {\calN} \int_{{\boldsymbol{q}}(t_1) =
{\boldsymbol{q}}_1}^{{\boldsymbol{q}}(t_2) = {\boldsymbol{q}}_2}
{\mathcal{D}}{\boldsymbol{q}} \ \prod_a \delta[
\dot{q}_a-f_a({\boldsymbol{q}})]\, , \label{eg.1.2}
\end{eqnarray}
 where
$ \delta[{\boldsymbol{f}} ]\equiv \prod_t  \delta
({\boldsymbol{f}}(t))$ is the functional version of Dirac's   $
\delta $-function. This result shows that quantization of the
system described by the Hamiltonian (\ref{eq.1.1}) retains its
deterministic character. The paths are squeezed onto the classical
trajectories   determined by the differential equations
 $\dot{q}_a = f_a({\boldsymbol{q}})$.
The time evolution amplitude (\ref{eg.1.2}) contains a sum  over
only the classical trajectories --- there are no quantum
fluctuations driving the system away from the classical paths,
which is precisely what we expect from a deterministic dynamics.

\vspace{3mm}

The amplitude (\ref{eg.1.2})
can be brought to a more intuitive form by  utilizing
 the identity
\begin{eqnarray}
\delta\left[ {\boldsymbol{f}}({{\boldsymbol{q}}})  -
\dot{{\boldsymbol{q}}} \right] \ = \ \delta[ {\boldsymbol{q}} -
{\boldsymbol{q}}_{\rm cl}]\ (\det {{M}})^{-1}\, ,
\end{eqnarray}
where ${M}$ is a functional matrix formed by the second derivatives
of the action $\overlinen \calS[{\boldsymbol{q}},\overlinen
{\boldsymbol{q}}]\equiv \int dt\,\overlinen L({\boldsymbol{q}},
\bar{{\boldsymbol{q}}}, \dot{{\boldsymbol{q}}},
\dot{\bar{{\boldsymbol{q}}}}) $\,:
\begin{eqnarray}
{{M}}_{ab}(t,t') \ = \ \left. \frac{\delta^2 \overlinen
\calS}{\delta q_a(t)\ \delta \overlinen{q}_b(t')}\
\right|_{{\boldsymbol{q}} = {\boldsymbol{q}}_{\rm cl}} \, .
\label{4.01}
\end{eqnarray}
The Morse index theorem then ensures that for sufficiently short
time intervals $t_2-t_1$ (before the system reaches its first focal
point), the classical solution with the initial condition
${{\boldsymbol{q}}}(t_1) = {\boldsymbol{q}}_1$ is unique. Note,
however, that because of the first-order character of the equations
of motion we are dealing with a Cauchy problem, which  may happen to
possess  no classical trajectory satisfying the two Dirichlet
boundary conditions ${{\boldsymbol{q}}}(t_1) = {\boldsymbol{q}}_1$,
${{\boldsymbol{q}}}(t_2) = {\boldsymbol{q}}_2$. If a trajectory
exists, Eq.~(\ref{eg.1.2}) can be brought to the form
\begin{eqnarray}
\langle {\boldsymbol{q}}_2,t_2| {\boldsymbol{q}}_1, t_1 \rangle
&=& {\bar\calN} \int_{{\boldsymbol{q}}(t_1) =
{\boldsymbol{q}}_1}^{{\boldsymbol{q}}(t_2) = {\boldsymbol{q}}_2}
{\mathcal{D}}{\boldsymbol{q}} \ \delta\left[{\boldsymbol{q}} -
{\boldsymbol{q}}_{\rm cl} \right]\, , \label{4.2}
\end{eqnarray}
where ${\bar\calN}\equiv {\calN}/(\det {M})$. We close this section
by observing that $\det M$ can be recast into more expedient form.
To do this we formally write
\begin{eqnarray}
\det M \ &=& \ \det\left|\!\left|  \left( \partial_t \delta_a^b +
\frac{\partial f_a({\boldsymbol{q}}(t))}{\partial q_b(t)}
\right)\delta(t-t') \right|\!\right| \ = \ \exp\left[
\mbox{Tr}\ln\left|\!\left|  \left( \partial_t \delta_a^b +
\frac{\partial f_a({\boldsymbol{q}}(t))}{\partial q_b(t)}
\right)\delta(t-t') \right|\!\right| \right]\nonumber \\
& = & \ \exp\left[ \mbox{Tr}\ln \partial_t \left|\!\left| \delta_a^b
\delta(t-t') + G(t-t')\frac{\partial
f_a({\boldsymbol{q}}(t'))}{\partial q_b(t')}\right|\!\right|
\right]\nonumber \\
&=& \ \exp\left[\mbox{Tr}(\ln \partial_t)\right]
\exp\left[\mbox{Tr}\ln \left|\!\left| \delta_a^b \delta(t-t') +
G(t-t')\frac{\partial f_a({\boldsymbol{q}}(t'))}{\partial
q_b(t')}\right|\!\right|\right]\, . \label{4.2.1}
\end{eqnarray}
Here $G(t-t')$ is the Green's function satisfying the equation
\begin{eqnarray*}
\partial_tG(t-t') \ = \ \delta(t-t')\, .
\end{eqnarray*}
Choosing $G(t-t') = \theta(t-t')$, and noting that the first factor
in Eq.(\ref{4.2.1}) is an irrelevant constant that can be
assimilated into ${\mathcal{N}}$ we have
\begin{eqnarray}
\det M \ &=&  \ \exp\left[\mbox{Tr} \ln \left|\!\left| \delta_a^b
\delta(t-t') + G(t-t')\frac{\partial
f_a({\boldsymbol{q}}(t'))}{\partial q_b(t')}\right|\!\right|\right]
\ = \ \exp\left[  \mbox{Tr}\left|\!\left|  \theta(t-t')
\frac{\partial f_a({\boldsymbol{q}}(t))}{\partial
q_b(t)}\right|\!\right| \right]\nonumber \\
&=& \ \exp\left[\frac{1}{2} \int_{t_1}^{t_2} \! dt \
{\boldsymbol{\nabla}}_{{\boldsymbol{q}}}
{\boldsymbol{f}}({\boldsymbol{q}}) \right]\, . \label{4.2.2}
\end{eqnarray}
In deriving Eq.(\ref{4.2.2}) we have used the fact that due to the
product of the $\theta$-function in the expansion of the logarithm,
 all terms
vanish but the first one. In evaluating the
generalized function
$\theta(x)$ at the origin
we have used the only consistent
midpoint rule~\cite{Pain}: $\theta(0) = 1/2$.
%
%
Using the  identity
\begin{eqnarray}
 \left.\exp\left[\frac{1}{2} \int_{t_1}^{t_2} \! dt \
{\boldsymbol{\nabla}}_{{\boldsymbol{q}}}
{\boldsymbol{f}}({\boldsymbol{q}}) \right] \right|_{{\boldsymbol{q}}
= {\boldsymbol{q}}_{\rm cl}}\ = \ \int {\mathcal{D}}
{\overline{{\boldsymbol{q}}}} \
\delta\left[\overline{{\boldsymbol{q}}} -
\overline{{\boldsymbol{q}}}_{\rm cl} \right] \
\exp\left[-\frac{1}{2} \int_{t_1}^{t_2} \! dt \
{\boldsymbol{\nabla}}_{\bar{\boldsymbol{q}}}
\dot{\bar{\boldsymbol{q}}} \right]\, ,
\end{eqnarray}
we can finally write the amplitude of transition in a suggestive
form
\begin{eqnarray}
\langle {\boldsymbol{q}}_2,t_2| {\boldsymbol{q}}_1, t_1 \rangle
&=& {\mathcal{N}} \int_{{\boldsymbol{q}}(t_1) =
{\boldsymbol{q}}_1}^{{\boldsymbol{q}}(t_2) = {\boldsymbol{q}}_2}
{\mathcal{D}}{\boldsymbol{q}}{\mathcal{D}}{\overline{\boldsymbol{q}}}
\ \delta[{{\boldsymbol{q}}} - {\boldsymbol{q}}_{\rm cl}]
\delta[\overline{{\boldsymbol{q}}} -
\overline{{\boldsymbol{q}}}_{\rm cl}] \ \exp\left[-\frac{1}{2}
\int_{t_1}^{t_2} \! dt \
{\boldsymbol{\nabla}}_{\bar{\boldsymbol{q}}}
\dot{\bar{\boldsymbol{q}}} \right] \nonumber \\
&=& {\mathcal{N}} \int_{{\boldsymbol{q}}(t_1) =
{\boldsymbol{q}}_1}^{{\boldsymbol{q}}(t_2) = {\boldsymbol{q}}_2}
{\mathcal{D}}{\boldsymbol{q}}{\mathcal{D}}{\overline{\boldsymbol{q}}}
\ \delta[{{\boldsymbol{q}}} - {\boldsymbol{q}}_{\rm cl}]
\delta[\overline{{\boldsymbol{q}}} -
\overline{{\boldsymbol{q}}}_{\rm cl}] \ \sqrt{\frac{\det K(t_2) }{
\det K(t_1)}}\, \ . \label{3.50}
\end{eqnarray}
Here $K(t)$ is the fundamental matrix of the solutions of the
system
\begin{eqnarray}
\dot{\bar{q}}_a = - \bar{q}_b \frac{\partial
f_b({\boldsymbol{q}})}{\partial q_a} \, .
\end{eqnarray}
$\det K(t)$ is then the corresponding Wronskian. Note that in the
particular case when ${\boldsymbol{\nabla}}_{{\boldsymbol{q}}}
{\boldsymbol{f}}({\boldsymbol{q}}) \equiv 0$, i.e., when the phase
flow preserves the volume of any domain in the {\em configuration}
space, the exponential in Eq.(\ref{3.50}) can be
dropped.\footnote{This corresponds to the situation when there are
no attractors in the configuration space
$\Gamma_{\boldsymbol{q}}.$} Because the exponent depends only on
the end points of $\bar{\boldsymbol{q}}$ variable it can be
removed by performing the trace over $\bar{{\boldsymbol{q}}}$.
%
%
%
As a result we can cast the quantum-mechanical partition function
(or generating functional) $Z$ into the form
\begin{eqnarray}
Z \ &=& \ {\mathcal{N}} \int
{\mathcal{D}}{\boldsymbol{q}}{\mathcal{D}}{\overline{\boldsymbol{q}}}
\ \delta[{{\boldsymbol{q}}} - {\boldsymbol{q}}_{\rm cl}]
\delta[\overline{{\boldsymbol{q}}} -
\overline{{\boldsymbol{q}}}_{\rm cl}] \  \exp\left[\int_{t_1}^{t_2}
[{\boldsymbol{J}}(t){\boldsymbol{q}}(t) +
\bar{{\boldsymbol{J}}}(t)\bar{{\boldsymbol{q}}}(t)]dt\right]\nonumber
\\
&=& \ {\mathcal{N}} \int {\mathcal{D}} q_a \ \delta[q_a - (q_a)_{\rm
cl}] \ \exp\left[\int^{t_2}_{t_1} dt \ J_a(t) q_a(t)  \right]\, .
\label{3.51}
\end{eqnarray}
Here the doubled vector notation $q_a = \{{\boldsymbol{q}},
\bar{\boldsymbol{q}}\}$ and $J_a \equiv \{{\boldsymbol{J}},
\bar{\boldsymbol{J}}\} $ was used.

\section{Path integral formulation of classical mechanics
- configuration-space approach}\label{SEc3}

Expressions (\ref{4.2}) and (\ref{3.51}) formally coincide with the
path-integral formulation of classical mechanics in configuration
space proposed by Gozzi~\cite{GozziI} and further developed by
Gozzi, Reuter, and Thacker~\cite{GozziII}(see also Ref.\cite{Elze2}
for recent applications). Let us briefly review aspects of this
which will be needed here. Consider the path-integral representation
of the generating functional of a quantum-mechanical system with
action $\calS[{\boldsymbol{q}}]$:
\begin{eqnarray}
{\calZ }_{\rm QM} = {\calN} \int {\mathcal{D}}{\boldsymbol{q}}\
e^{-i \calS[{\boldsymbol{q}}]/\hbar } \exp\left[\int
{\boldsymbol{J}}(t){\boldsymbol{q}}(t)dt\right]\, . \label{@genf}
\label{4.0}
\end{eqnarray}
We assume in this context that there are no constraints that would
make the measure more complicated as in Eq.~(\ref{frad}). Gozzi {\em
et al.} proposed to describe classical mechanics by a generating
functional of the form (\ref{@genf}) with an obviously  modified
integration measure which gives  equal weight to all classical
trajectories and zero weight to all others
\begin{eqnarray}
{\calZ }_{\rm CM} = \tilde{{\calN}} \int
{\mathcal{D}}{\boldsymbol{q}} \ \delta[{\boldsymbol{q}}-
{\boldsymbol{q}}_{\rm cl}]
 \exp\left[\int
{\boldsymbol{J}}(t){\boldsymbol{q}}(t)dt\right]\, . \label{4.3}
\end{eqnarray}
Although the form of the partition function (\ref{4.3}) is not
derived but {\em postulated}, we show in Appendix B that it can be
heuristically understood either as the ``classical" limit of the
stochastic-quantization partition function (c.f., Appendix BI), or
as a results of the classical limit of the closed-time path integral
for the transition probability of systems coupled to a heat bath
(c.f., Appendix BII). This, in turn, indicates that it would be
formally more correct to associate (\ref{4.3}) with the {\em
probability} of transition or (via the stochastic-quantization
passage) with the {\em Euclidean} amplitude of
transition~\cite{Zinn-JustinII}. Albeit (\ref{4.3}) cannot be
generally obtained from (\ref{4.0}) by a semiclassical limit {\em
\`{a} la} WKB (which can be recognized by the absence of a phase
factor $\exp(i/\hbar {\mathcal{A}}(q_{\rm cl}))$ in (\ref{4.3})) it
may happen that even ordinary amplitudes of transition posses this
form. This is the case, for instance, when the number of degrees of
freedom is doubled or when one deals with closed-time-path
formulation of thermal quantum theory. Yet, whatever is the origin
or motivation for (\ref{4.3}), it will be its formal structure and
mathematical implications that will interest us here most.

\vspace{3mm}


\vspace{3mm}

To proceed we note that an alternative way of writing (\ref{4.3})
is
\begin{eqnarray}
{\calZ }_{\rm CM}\ = \ \tilde{{\calN}} \int
{\mathcal{D}}{\boldsymbol{q}} \ \delta\left[ \frac{\delta
\calS}{\delta {\boldsymbol{q}}} \right] \ \det \left| \frac{\delta^2
\calS }{\delta q_a (t) \ \delta q_b(t')} \right| \ \exp\left[\int
{\boldsymbol{J}}(t){\boldsymbol{q}}(t)dt\right]\, . \label{4.4}
\end{eqnarray}
By representing the $\delta$ functional in the usual way as a
functional Fourier integral,
\begin{eqnarray}
\delta\left[ \frac{\delta \calS}{\delta {\boldsymbol{q}}} \right] =
\int {\mathcal{D}} {\mathbf{\lambda}} \ \exp\left( i
\int_{t_1}^{t_2} dt \ {\boldsymbol{\lambda}}(t) \frac{\delta
\calS}{\delta {\boldsymbol{q}}(t)} \right)\, ,
\end{eqnarray}
and the functional determinant as a functional integral over two
real time-dependent Grassmannian {\em ghost variables\/} $c_a(t)$
and $\overlinen{c}_a(t)$,
\begin{eqnarray}
\det \left| \frac{\delta^2 \calS }{\delta q_a (t) \ \delta q_b(t')}
\right| = \int {\mathcal{D}}{\boldsymbol{c}} {\mathcal{D}}
\overlinen{{\boldsymbol{c}}} \ \exp\left[ \int_{t_1}^{t_2}dt
\int_{t_1}^{t_2} dt' \ \overlinen{c}_a(t) \frac{\delta^2 \calS
}{\delta{q_a} (t) \ \delta{q_b}(t')} \ {c_b}(t')\right]\, ,
\end{eqnarray}
we obtain
\begin{eqnarray}
{\calZ }_{\rm CM} \ = \ \int
{\mathcal{D}}{\boldsymbol{q}}{\mathcal{D}}{\boldsymbol{\lambda}}{\mathcal{D}}{\boldsymbol{c}}
{\mathcal{D}}\overlinen{{\boldsymbol{c}}} \ \exp\left[ i
{\mathcal{S}} + \int_{t_1}^{t_2} dt \
{{\boldsymbol{J}}}(t){{\boldsymbol{q}}}(t) \right]\, ,
\label{4.5a}
\end{eqnarray}
with the new action
\begin{eqnarray}
{\mathcal{S}}[{\boldsymbol{q}}, \overlinen{{\boldsymbol{c}}},
{\boldsymbol{c}}, {\boldsymbol{\lambda}}] \equiv \
\int_{t_1}^{t_2} dt \ {\boldsymbol{\lambda}}(t)\frac{\delta
\calS}{\delta {\boldsymbol{q}}(t)} - i\int_{t_1}^{t_2}dt
\int_{t_1}^{t_2} dt' \ \overlinen{c}_a(t) \frac{\delta^2 \calS
}{\delta{q_a} (t) \ \delta{q_b}(t')} \ {c_b}(t')\, . \label{4.5}
\end{eqnarray}
Since ${\calZ }_{\rm CM}$ together with the action (\ref{4.5})
formally result from the classical limit of the
stochastic-quantization partition function, it comes as no
surprise that ${\mathcal{S}}$ exhibits BRST (and anti-BRST)
supersymmetry. It is simple to check that $\mathcal{S}$ does not
change under  the supersymmetry transformations
\begin{eqnarray}
\delta_{{\rm BRST\,}}  {\boldsymbol{q}} = \overlinen{\varepsilon}
{\boldsymbol{c}}\,, \;\; \delta_{{\rm BRST\,}}  {\boldsymbol{c}} = 0\, , \;\;
\delta_{{\rm BRST\,}}  \overlinen{{\boldsymbol{c}}} = -i\overlinen{\varepsilon}
{\boldsymbol{\lambda}}\, , \;\; \delta_{{\rm BRST\,}}  {\boldsymbol{\lambda}} =
0\, , \label{4.6}
\end{eqnarray}
where $\overlinen{\varepsilon}$ is a Grassmann-valued parameter
(the corresponding anti-BRST transformations are related with
(\ref{4.6}) by charge conjugation). Indeed, the variations of the
two terms in (\ref{4.5}) read
\begin{eqnarray}
&&\delta_{{\rm BRST\,}} \left[\int_{t_1}^{t_2} dt \
{\boldsymbol{\lambda}}(t)\frac{\delta \calS}{\delta
{\boldsymbol{q}}(t)} \right] \ = \
\overlinen{\varepsilon}\int_{t_1}^{t_2} dt \int_{t_1}^{t_2} dt' \
\lambda_a(t) \frac{\delta^2 \calS}{\delta q_a(t) \delta q_b(t')} \
c_b(t')\,
,\label{4.7} \\
&&~\nonumber \\
&&\delta_{{\rm BRST\,}} \left[  \int_{t_1}^{t_2}dt \int_{t_1}^{t_2} dt' \
{\overlinen{c}}_a(t) \frac{\delta^2 \calS}{\delta q_a(t) \delta
q_b(t')} \ c_b(t') \right] \ = \ -i \overlinen{\varepsilon}
\int_{t_1}^{t_2}dt \int_{t_1}^{t_2} dt' \ \lambda_a(t)
\frac{\delta^2 \calS}{\delta q_a(t) \delta q_b(t')} \ c_b(t')\nonumber
\\
&& ~\nonumber \\
&&\mbox{\hspace{15mm}} + \ \int_{t_1}^{t_2}dt \int_{t_1}^{t_2} dt'
\int_{t_1}^{t_2}dt'' \ {\overlinen{c}}_a(t) \frac{\delta^3
\calS}{\delta q_a(t) \delta q_b(t') \delta q_c(t'')} \
\overlinen{\varepsilon} \ c_c(t'') c_b(t')\, . \label{4.8}
\end{eqnarray}
\\
\noi The second term on the RHS of (\ref{4.8}) vanishes because the
functional derivative of $\calS$ is symmetric in $c\leftrightarrow
b$ whereas the term $c_c c_b$ is anti-symmetric. Inserting
Eqs.(\ref{4.7}) and (\ref{4.8}) into the action we clearly find
$\delta_{{\rm BRST\,}}{\mathcal{S}} = 0$. As noted
in~\cite{GozziII}, the ghost fields $\overlinen{{\boldsymbol{c}}}$
and ${\boldsymbol{c}}$ are mandatory at the classical level as their
r\^{o}le is to cut off the fluctuations {\em perpendicular\/} to the
classical trajectories. On the formal side,
$\overlinen{{\boldsymbol{c}}}$ and ${\boldsymbol{c}}$ may be
identified with Jacobi fields~\cite{GozziII,DeWitt}. The
corresponding BRST charges
are related to Poincar\'{e}-Cartan integral
invariants~\cite{GozziIII}.

\vspace{3mm}

By analogy with the stochastic quantization the path integral
(\ref{4.5a}) can, of course, be rewritten in a compact form with
the help of a superfield~\cite{GozziI,Zinn-JustinII}
\begin{eqnarray}
\Phi_a(t, \theta, \overlinen{\theta}) \ = \ q_a(t) + i\theta c_a(t)
-i\overlinen{\theta} \overlinen{c}_a(t) + i \overlinen{\theta}\theta
\lambda_a(t)\, , \label{3.23}
\end{eqnarray}
in which $\theta$ and $\overlinen{\theta}$ are anticommuting
coordinates extending the configuration space of $q_a$ variable to a
superspace. The latter is nothing but the degenerate case of
supersymmetric field theory in $d=1$ in the superspace formalism of
Salam and Strathdee~\cite{SS1}. In terms of superspace variables we
see that
\begin{eqnarray}
\int d\overlinen{\theta} d\theta \ {\calS}[{\boldsymbol{\Phi}}] &=&
\int dt d\overlinen{\theta} d\theta \ L({\boldsymbol{q}}(t) +
i\theta {\boldsymbol{c}}(t) - i \overlinen{\theta}
\overlinen{\boldsymbol{c}}(t) + i
\overlinen{\theta}\theta \boldsymbol{\lambda}(t) )\nonumber \\
&=& \int d\overlinen{\theta} d\theta \ {\calS}[{\boldsymbol{q}}] + \int
dt d\overlinen{\theta} d\theta \ \left( i\theta {\boldsymbol{c}}(t)
- i \overlinen{\theta} \overlinen{\boldsymbol{c}}(t) +
i\overlinen{\theta}\theta \boldsymbol{\lambda}\right) \frac{\delta
{\calS}}{\delta
\boldsymbol{q}(t)}\nonumber \\
&& + \ \int dt dt' d\overlinen{\theta} d\theta \ \theta
c_a(t)\frac{\delta^2 {\calS}}{\delta q_a(t) \delta q_b(t')} \
\overlinen{\theta} \overlinen{c}(t').
\end{eqnarray}
Using the standard integration rules for Grassmann variables, this
becomes equal to $-i{\mathcal{S}}$. Together with the identity
${\mathcal{D}} {\boldsymbol{\Phi}} =
{\mathcal{D}}{\boldsymbol{q}}{\mathcal{D}}
{\boldsymbol{c}}{\mathcal{D}}\overlinen{{\boldsymbol{c}}}
{\mathcal{D}}{\boldsymbol{\lambda}}$ we may therefore express the
classical partition functions (\ref{4.3}) and (\ref{4.4}) as a
supersymmetric path integral with fully fluctuating paths in
superspace,
\begin{eqnarray}
{\calZ }_{\rm CM} \ = \ \int {\mathcal{D}} {\boldsymbol{\Phi}} \
\exp\left\{- \int d\theta d\overlinen{\theta} \
\calS[{\boldsymbol{\Phi}}](\theta, \overlinen{\theta}) + \int dt
d\theta d\overlinen{\theta} \ {\boldsymbol{\Gamma}}(t, \theta,
\overlinen{\theta}){\boldsymbol{\Phi}}(t, \theta,
\overlinen{\theta})\right\}\, . \label{3.24}
\end{eqnarray}
Here we have defined the supercurrent ${\boldsymbol{\Gamma}}(t,
\theta, \overlinen{\theta}) = \overlinen{\theta} \theta
{\boldsymbol{J}}(t)$.

\vspace{3mm}

It is interesting to find the most general form of an action $\calS$
for which the classical path integral (\ref{3.24}) coincides with
the quantum-mechanical path integral of the system, or, in other
words, for which a theory would possess at the same time
deterministic and quantal character. As already mentioned, the
Grassmannnian ghost variables are responsible for the
deterministic nature of the partition function. It is obvious that
if the ghost sector could somehow be factored out we would extend
the path integration to all fluctuating paths in
${\boldsymbol{q}}$-space. By formally writing
\begin{eqnarray}
 \frac{\delta^2 \calS }{\delta{q_k} (t) \ \delta{q_l}(t')} \ = \
 {\mathcal{F}}_{kl}\left( t,t', q_m, \frac{\delta \calS}{\delta q_n}\right)\, ,
 \;\;\;\;\;  k,l,m,n = 1, \ldots, N\, ,
\label{4.9}
\end{eqnarray}
we see that the factorization will occur if and only if the
(distribution valued) functional ${\mathcal{F}}_{kl}(\ldots)$ is
$q_m$ independent when evaluated on shell, i.e.,
${\mathcal{F}}_{kl}(t,t', q_m, 0 ) = F_{kl}(t,t')$. This is a simple
consequence of Eq.(\ref{4.4}) where the determinant is factorizable
if and only if it is ${\boldsymbol{q}}$-independent at $\delta
\calS/\delta {{\boldsymbol{q}}} = 0$.


\vspace{3mm}

In order to provide a correct Feynman weight to every path we must,
in addition, identify
\begin{eqnarray}
\calS[{\boldsymbol{q}}] = \int_{t_1}^{t_2} dt \ \lambda_m
\frac{\delta \calS[{\boldsymbol{q}}]}{\delta q_m}
\, , \label{3.26}
\end{eqnarray}
as can be seen from (\ref{4.5}) after factoring out the second term.
Assuming that $L = L(q_l, \dot{q_l})$ (i.e., a scleronomic system)
and that the Hessian is regular,
 the condition (\ref{3.26}) shows that
$\lambda_k = \lambda_k(q_l, \dot{q_k})$. In addition,
it is obvious
on
dimensional grounds
 that $\left[ \lambda_l  \right] =
\left[ q_l \right]$. This, in turn, implies that $\lambda_k  =
\alpha_{kl}q_l$, where $\alpha_{lk}$ is some real ($t$-independent)
matrix. To determine the latter we functionally expand $ \calS$ in
(\ref{3.26}) around $q_k$ and compare both sides. The resulting
integrability condition reads:
\begin{eqnarray}
\left(\delta_{ji} -\alpha_{ji}\right)\frac{\delta \calS}{\delta
q_{j}(t)}\ \delta (t-t') \ = \ \alpha_{l\!j} \ q_j(t) \
\frac{\delta^2  \calS}{\delta q_l(t) \delta q_i(t')}\, ,
\label{3.28}
\end{eqnarray}
which  is evidently compatible with the condition (\ref{4.9}).
When $\alpha_{ij}$ is diagonalizable we can pass to a polar basis
and write (\ref{3.26}) in more manageable form, namely
\begin{eqnarray}
\calS[{\boldsymbol{q}}]\ = \ \int_{t_1}^{t_2} dt \ \sum_i \alpha_i
q_i(t) \frac{\delta \calS[{\boldsymbol{q}}]}{\delta q_i(t)}\, .
\label{3.33}
\end{eqnarray}
For simplicity, we do not use new symbols
 for
transformed ${\boldsymbol{q}}$'s.

\vspace{3mm}

To proceed we  assume that the kinetic energy is  quadratic in
${\boldsymbol{q}}$ and $\dot{{\boldsymbol{q}}}$. Then
Eq.(\ref{3.33}) implies that $L_{\rm kin}$ must be liner in
$\dot{{\boldsymbol{q}}}$. As such, one can always write (modulo the
total derivative)
\begin{eqnarray}   \label{@withB}
L_{\rm kin} = \sum_{i,j} {\mB}_{ij} \ q_i(t) \dot{q}_j(t)\, ,
\end{eqnarray}
with ${\mB}$ being an upper triangular matrix. Comparing $L_{\rm
kin}$ on both sides of (\ref{3.33}) we arrive at the equation
\begin{eqnarray}
(\alpha_m - 1) {\mB}_{im} = {\mB}_{mi} \alpha_m \,
\, \Rightarrow  \, \, (\mB -
\mB^{\top}){\boldsymbol{\alpha}} = \mB\, ,
\label{3.37}
\end{eqnarray}
with no Einstein's summation convention applied here. Because ${\mB}$ is
upper triangular, the first part of  Eq.(\ref{3.37}) implies that
the only eigenvalues of $\alpha_{ij}$ are $1$ and $0$. Thus,
${\boldsymbol{\alpha}}$ can be reduced to the block form
\begin{eqnarray}
{\boldsymbol{\alpha}} \ = \ \left[ \begin{tabular}{c|c} 0 & 0 \\
\hline 0 & \ide
\end{tabular}
\right]\, , \label{3.38}
\end{eqnarray}
where $\ide$ is a $r\times r$ ($r\leq N$) unit matrix.
 Using the equation $({\mB} -
{\mB}^{\top}){\boldsymbol{\alpha}} = {\mB} $ we see that
${\mB}$   has the
 block structure
\begin{eqnarray}
\mbox{${\mB}$} \ = \
\left[
\begin{tabular}{c|c} 0 &
${\mB}_2$
\\
\hline
0 & 0 \\
\end{tabular}
\right] \, . \label{3.39}
\end{eqnarray}
where ${\mB}_2$ is an $(N-r) \times r$ matrix. To determine $r$ we
use the fact that ${\boldsymbol{\alpha}}$ is idempotent, i.e.,
 ${\boldsymbol{\alpha}}^2 =
{\boldsymbol{\alpha}}$. Multiplying
$({\mB} -
{\mB}^{\top}){\boldsymbol{\alpha}} = {\mB} $ by
$\boldsymbol{\alpha}$ we find
\begin{eqnarray}
\begin{array}{ll}
{\mB}{\boldsymbol{\alpha}} = {\mB}\, ,~~~~~
{\mB}^\top {\boldsymbol{\alpha}} = 0 \, .
\end{array}
\end{eqnarray}
From ${\mB}{\boldsymbol{\alpha}} = {\mB}$ follows that rank$({\mB})=
{\mbox{rank}}({\boldsymbol{\alpha}})=r$, whereas ${\mB}^\top(\ide -
{\boldsymbol{\alpha}}) = {\mB}^\top$ implies that rank$({\mB}^\top)=
{\mbox{rank}}(\ide - {\boldsymbol{\alpha}})$. Utilizing the identity
${\mbox{rank}}({\mB}) = {\mbox{rank}}({\mB}^\top)$ we derive
 $r = {\mbox{rank}}({\boldsymbol{\alpha}})
= {\mbox{rank}}(\ide - {\boldsymbol{\alpha}}) = (N-r)$, and thus $r
= N/2$. Thus the condition (\ref{3.33}) can be satisfied only for an
even number $N$ of degrees of freedom. An immediate further
consequence of (\ref{3.39}) is   that we can rewrite (\ref{@withB})
as
\begin{eqnarray}
L_{\rm kin} = \sum_{i,j =1}^{N/2} {\mB}_{i,(N/2+j)} \ \dot{q}_i
q_{N/2 +j} \, . \label{3.36}
\end{eqnarray}
Denoting $\alpha_{N/2 + i}$, $q_{N/2 + i}$ and $\lambda_{N/2 +i}$
($i = 1,\ldots,N/2$) as ${\bar{\alpha}}_{i}$, $\bar{q}_i$, and
$\bar{\lambda}_{i}$, respectively [hence, ${\boldsymbol{\lambda}}
= {\boldsymbol{0}}$ and $\bar{{\boldsymbol{\lambda}}} =
\bar{{\boldsymbol{q}}}~$], then Eq.(\ref{3.33}) reads
\begin{eqnarray}
\bar{\calS}[{\boldsymbol{q}}, \bar{{\boldsymbol{q}}}] =
\int_{t_1}^{t_2} dt \ \bar{{\boldsymbol{q}}}(t) \frac{\delta
\bar{\calS}[{\boldsymbol{q}}, \bar{{\boldsymbol{q}}}]}{\delta
\bar{{\boldsymbol{q}}}(t) }\, . \label{3.35}
\end{eqnarray}
%
Here $\bar{\calS}[{\boldsymbol{q}}, \bar{{\boldsymbol{q}}}] =
{\calS}[q_1, \ldots, q_{2N}]$. The result (\ref{3.35}) can be
obtained also in a different way. Indeed, in Appendix C we show that
(\ref{3.33}) is a so-called Euler-like functional
\begin{eqnarray}
\calS[{\boldsymbol{q}}] = \int_{t_1}^{t_2} dt \ r(t) L\!\left(
r^{-\alpha_1}(t)q_1(t), \ldots,r^{-\alpha_N}(t)q_N(t),
\frac{d\left(r^{-\alpha_1}(t)q_1(t)\right)}{dt}, \ldots,
\frac{d\left(r^{-\alpha_N}(t)q_N(t)\right)}{dt}
 \right)
%
 \, ,\label{3.31}
\end{eqnarray}
with $r(t)$ being an arbitrary function of $q_k$ whose variations
vanish at the ends $\delta r(t_i) = \delta r(t_f) = 0$ if all $
\delta q_k$'s have this property.
 In particular, we may
 chose $r$ to be any finite power $ q_k^{1/\alpha_k}$
 (for $k = 1, \ldots, N$), in which case
\begin{eqnarray}
\calS[{\boldsymbol{q}}] = \int_{t_1}^{t_2} dt \ q_k^{1/\alpha_k}
L\!\left( \frac{q_1}{q_k^{\alpha_1/\alpha_k}},
 \dots, \stackrel{\stackrel{\mbox{\footnotesize{$k$}}}{\downarrow}}{1},
 \dots, \frac{q_N}{q_k^{\alpha_N/\alpha_k}},
 \frac{d\left(q_1/q_k^{\alpha_1/\alpha_k}\right)}{dt}, \ldots,
 \stackrel{\stackrel{\mbox{\footnotesize{$k$}}}{\downarrow}}{0}, \dots,
 \frac{d\left(q_N/q_k^{\alpha_N/\alpha_k}\right)}{dt}
 \right).
 \label{3.30}
\end{eqnarray}
Assuming, as before, that the kinetic term in $L$ is quadratic  in
${\boldsymbol{q}}$ and $\dot{\boldsymbol{q}}$, we arrive at
${\boldsymbol{\alpha}}$ as in (\ref{3.38}), and the action
(\ref{3.30}) reduces again to (\ref{3.35}).

\vspace{3mm}

One can incorporate the constraints on $\alpha_i$ (or $\lambda_i$)
by inserting a corresponding  $\delta$-functional into the path
integral (\ref{4.5a}). This leads to the most general generating
functional with the above-stated property:
\begin{eqnarray}
{\calZ }_{\rm CM} &=&  \int
{\mathcal{D}}{{\boldsymbol{q}}}{\mathcal{D}}{\overlinen{{\boldsymbol{q}}}}
{\mathcal{D}}{{\boldsymbol{\lambda}}}
{\mathcal{D}}{\overlinen{{\boldsymbol{\lambda}}}} \
\delta[{\boldsymbol{\lambda}}]
\delta[\overlinen{{\boldsymbol{\lambda}}} -
\overlinen{{\boldsymbol{q}}}] \ \exp\!\left[i \!\!\int_{t_1}^{t_2}
dt \ {\boldsymbol{\lambda}}\frac{\delta \overlinen
\calS[{\boldsymbol{q}}, \overlinen{{\boldsymbol{q}}}]}{\delta
{\boldsymbol{q}} } + i \!\!\int_{t_1}^{t_2} dt \
\overlinen{{\boldsymbol{\lambda}}}\ \frac{\delta \overlinen
\calS[{\boldsymbol{q}}, \overlinen{{\boldsymbol{q}}}]}{\delta
\overlinen{{\boldsymbol{q}}} }  + \int_{t_1}^{t_2} dt \sum_{k=1}^N\! J_k q_k \right]\nonumber \\
&~& \nonumber \\
&=& \int
{\mathcal{D}}{{\boldsymbol{q}}}{\mathcal{D}}{\overlinen{{\boldsymbol{q}}}}
\ \exp\!\left[ i\!\!\int_{t_1}^{t_2} dt \
\overlinen{{\boldsymbol{q}}}\ \frac{\delta \overlinen
\calS[{\boldsymbol{q}}, \overlinen{{\boldsymbol{q}}}]}{\delta
\overlinen{{\boldsymbol{q}}} }
+  \int_{t_1}^{t_2} dt \sum_{k=1}^N J_k q_k \right] \nonumber \\
&~& \nonumber \\
&=& \int
{\mathcal{D}}{{\boldsymbol{q}}}{\mathcal{D}}{\overlinen{{\boldsymbol{q}}}}
\ \exp\!\left[ i \!\!\int_{t_1}^{t_2} dt \,\overlinen L + \int dt
\sum_{k=1}^N J_k q_k \right]\, .
\label{3.27}
\end{eqnarray}
An irrelevant
 normalization factor has been dropped.
 The Lagrangian $\overlinen L$ coincides precisely
with the Lagrangian (\ref{lag1}), and describes therefore
't\,Hooft's deterministic system. Hence within the above assumptions
there are no other systems with the peculiar property that their
full quantum properties are classical. Among other things, the
latter also indicates that the Koopman-von~Neumann operatorial
formulation of classical mechanics~\cite{KN1} when applied to
't~Hooft systems must agree with its canonically quantized
counterpart.




\section{'t\,Hooft's information loss as a first-class primary
constraint}\label{SEc4}

As observed in Section~II, the Hamiltonian (\ref{eq.1.1}) is not
bounded from below, and this is true for any function $f_i$. Thus,
no deterministic system with dynamical equations $\dot{q}_i =
f_i({\boldsymbol{q}})$ can describe a  physically acceptable {\em
quantum world\/}. Its Hamiltonian would not be stable and we could
build a perpetuum mobile. To deal with this problem we will employ
't\,Hooft's procedure~\cite{tHooft}. We
 assume that the system (\ref{eq.1.1}) has $n$
conserved, irreducible charges $C_i$, i.e.,
\begin{eqnarray}
\{ C_i, H \} = 0\, , \;\;\;\; i = 1, \ldots, n\, .      \label{@char}
\end{eqnarray}
In order to enforce a lower bound upon $H$,
 't\,Hooft split the Hamiltonian
 as
 $H = H_+ - H_-$ with  both $H_+$ and $H_-$
having lower bounds. Then  he imposed the condition that $H_-$
should be zero on the physically accessible part of phase space,
i.e.,
\begin{eqnarray}
H_- \ \approx \ 0\, . \label{4.1}
\end{eqnarray}
This will make the actual dynamics
 governed by the reduced Hamiltonian
$H_+$ which is
bounded from below,
 by definition.

\vspace{3mm}

To ensure that the above splitting is conserved in time one must
require that $\{ H_-, H \} = \{ H_+, H \} = 0$. The latter is
equivalent to the statement that $\{ H_+, H_- \} = 0$. Since
the
charges
$C_i$ in (\ref{@char}) form an irreducible set,
 the Hamiltonians
 $H_+$
and $H_-$
 must
be
 functions of the charges and $H$:
$H_+ = F_+(C_k,H)$ and $H_- = F_-(C_k,H)$.
 There is a certain amount of
flexibility in finding $F_-$ and $F_+$, but for convenience's sake
we confine ourselves to the following choice
\begin{eqnarray}
H_+ \ = \ \frac{[H + \sum_i a_i(t) C_i]^2}{4 \sum_i a_i(t) C_i} \,
, \; \; H_- \ = \ \frac{[H - \sum_i a_i(t) C_i]^2}{4 \sum_i a_i(t)
C_i} \, , \label{FCH}
\end{eqnarray}
where $a_i(t)$ are independent of  ${\boldsymbol{q}}$ and
${\boldsymbol{p}}$ and will be specified later. The lower bound is
then achieved by choosing $\sum_i a_i(t) C_i$ to be positive
definite. In the following it will also be important to select the
combination of $C_i$'s in such a way that it depends solely on
${\boldsymbol{q}}$ (this condition may not necessarily be achievable
for general $f_a({\boldsymbol{q}})$). Thus, by imposing $H_- \approx
0$ we obtain the weak reduced Hamiltonian $H \approx H_+ \approx
\sum_i a_i(t)C_i$.

\vspace{3mm}

The constraint (\ref{4.1}) (resp (\ref{FCH})) can be motivated by
dissipation or information loss~\cite{tHooft3,BJV1,BMM1}. In
Appendix D we show that the {\em explicit} constraint (\ref{4.1})
does not generate any new (i.e., secondary) constraints when added
to the existing constraints (\ref{2.5}). In addition, this new set
of constraints corresponds to $2N$ second-class constraints and {\em
one} first-class constraint (see also Appendix D).
It is well known in the theory of constrained systems that the
existence of first-class constraints signals the presence of a gauge
freedom in Hamiltonian theory. This is so because the Lagrange
multipliers affiliated with first-class constraints cannot be fixed
from dynamical equations alone~\cite{Dir}. The time evolution of
observable (physical) quantities, however, cannot be affected by the
arbitrariness in Lagrange multipliers. To remove this superfluous
freedom that is left in the formalism we must pick up a gauge, i.e.,
impose a set of conditions that will eliminate the above redundancy
from the description. It is easy to see that the number of
independent gauge conditions must match the number of first-class
constraints. Indeed, the requirement on a physical quantity (say
$f$) to have a unique time evolution on the constraint submanifold
${\mathcal{M}}$, i.e.,
\begin{eqnarray}
\dot{f} \ \approx \ \{ f, \bar{H}\} \ + \ \sum_{i=1}^{m}v_i \{ f,
\varphi_i\} \ + \  \sum_{k=1}^{m'}u_k \{ f, \phi_k \}\, ,
\end{eqnarray}
implies that
\begin{eqnarray}
\{ f, \varphi_i\} \ \approx \ 0\, . \label{4.24}
\end{eqnarray}
The constraints $\varphi_i$ and $\phi_k$ represent first and
second-class constraints, respectively. First-class constraints
have, by definition, weakly vanishing Poisson's brackets with all
other constraints; any other constraint that is not first class is
second-class. While the Lagrange multipliers $u_k$ can be uniquely
fixed from the dynamics by consistency conditions (c.f. Appendices A
and D) this cannot be done for the $v_i$'s. In this way (\ref{4.24})
represents an obligatory condition for a quantity $f$ to be
observable. Equation (\ref{4.24}) can be considered as a set of $m$
first-order differential equations on the constrained surface with
the relation $\{\varphi_i, \varphi_j \} \ \approx \ 0$ serving as
the integrability condition~\cite{Dir,Sunder}. Thus, $f$ is uniquely
defined by its values on the submanifold of the initial conditions
for Eq.(\ref{4.24}). As a result, the above initial value surface
describes the true degrees of freedom. By denoting the dimension of
the constraint manifold as $D$ we see that the dimension of the
submanifold of initial conditions must be $D-m$. We can take this
submanifold to be a surface $\Gamma^*$ specified by the equations
\begin{eqnarray}
\varphi_i \ &=& \ 0\, , \;\;\;\;\;\; i = 1, \ldots, m\, ,\nonumber \\
\phi_k \ &=& \ 0 \, , \;\;\;\;\;\; k = 1, \ldots, m'\, ,\nonumber \\
\chi_l \ &=& \ 0 \, , \;\;\;\;\;\; l = 1, \ldots, m\, .
\label{4.24b}
\end{eqnarray}
The $m$ subsidiary conditions $\chi_l$ are the sought gauge
constraints. The functions $\chi_l$ must clearly satisfy the
condition
\begin{eqnarray}
\det\| \{\chi_l , \varphi_i \} \| \ \neq \ 0\, , \label{4.25}
\end{eqnarray}
as only in such a case we can determine specific values for the
multipliers $v_i$ from the dynamical equation for $\chi_l$ (this is
because the time derivative of any constraint, and hence also
$\chi_l$, must be zero). Therefore only when the condition
(\ref{4.25}) is satisfied do the constraints (\ref{4.24b}) indeed
describe the surface of the initial conditions.

\vspace{3mm}

The preceding discussion implies that in our case the surface
$\Gamma^*$ is defined by
\begin{eqnarray}
\varphi({\boldsymbol{q}},\bar{\boldsymbol{q}},
{\boldsymbol{p}},\bar{\boldsymbol{p}}) \ &=& \ 0 \, ,
\mbox{\hspace{0.6cm}} \chi({\boldsymbol{q}},\bar{\boldsymbol{q}},
{\boldsymbol{p}},\bar{\boldsymbol{p}}) \ = \ 0 \, , \label{4.26}
\\
\mbox{\hspace{1cm}} \phi_i({\boldsymbol{q}},\bar{\boldsymbol{q}},
{\boldsymbol{p}},\bar{\boldsymbol{p}}) \ &=& \ 0 \, , \;\;\;\;\;\;
i = 1, \ldots, 2N\, .
\end{eqnarray}
The explicit form of $\varphi$ is found in Appendix~D where we
show that $\varphi \approx H - \sum a_i C_i$.
Apart from condition (\ref{4.25}) we shall further restrict our
choice of $\chi$ to functions satisfying the simultaneous equations
\begin{eqnarray}
\{ \chi, \phi_i \} \ = \ 0\, , \;\;\;\;\; i = 1, \ldots, 2N\, .
\label{4.27}
\end{eqnarray}
Such a choice is always possible (at least in a weak
sense)~\cite{Senj} and it will prove crucial in the following.



\vspace{3mm}

In order to proceed further we begin by reexamining
Eq.(\ref{3.27}). The latter basically states that
\begin{eqnarray}
Z_{\rm CM} \ = \ \int {\mathcal{D}}{\boldsymbol{q}} \
\delta\!\left[{\boldsymbol{q}} - {\boldsymbol{q}}_{c}
\right] \ \exp\left[
\int_{t_1}^{t_2} dt \ {\boldsymbol{q}}(t){\boldsymbol{J}}(t)
\right]\, . \label{3.40}
\end{eqnarray}
We may now formally invert the steps leading to Eq.(\ref{eg.1.2}),
i.e., we introduce auxiliary momentum integrations and go over to
the canonical representation of (\ref{3.40}).  Correspondingly
Eq.(\ref{3.40}) can be recast into
\begin{eqnarray*}
Z_{\rm CM} =  \int
{\mathcal{D}}{\boldsymbol{p}}{\mathcal{D}}{\boldsymbol{q}}{\mathcal{D}}
\bar{\boldsymbol{p}}{\mathcal{D}}\bar{\boldsymbol{q}}
\sqrt{\left|\det\|\{ \phi_i, \phi_j \}\| \right|}
\prod_{i=1}^{2N} \delta[\phi_i]  \exp\left[ i\! \int_{t_1}^{t_2}
\!dt \,[{\boldsymbol{p}}\dot{\boldsymbol{q}} +
\bar{\boldsymbol{p}}\dot{\bar{\boldsymbol{q}}} - H] +
\int_{t_1}^{t_2}\! dt\, [{\boldsymbol{q}}{\boldsymbol{J}} +
\bar{\boldsymbol{q}}\bar{\boldsymbol{J}}] \right] .
\end{eqnarray*}
Due to $\delta$-functions in the integration we could substitute
't~Hooft's Hamiltonian $H$ for the canonical Hamiltonian $\bar{H}$.
It should be stressed that despite its formal appearance and the
phase-space disguise, the latter is still the classical partition
function {\em{\`{a}} la } Gozzi {\em et al.}.

\vspace{3mm}

To include the constraints (\ref{4.26}) into
(\ref{3.27}) we must be a bit cautious. A na\"{\i}ve intuition would
dictate that the functional $\delta$ functions $\delta[\chi]$ and
$\delta[\varphi]$ should be inserted into the path-integral measure
for $Z_{\rm CM}$. This would be, however, too simplistic as a mere
inclusion of $\delta$ functions into $Z_{\rm CM}$ would not
guarantee that the physical content of the theory that resides in
the generating functional $Z_{\rm CM}$ is independent of the choice
$\chi$. Indeed, utilizing the fact that the generators of gauge
transformations are the first class constraints~\cite{Sunder} we can
write that
\begin{eqnarray}
\delta \chi \ = \ \varepsilon \{\chi , \varphi \} + C \varphi \
\approx \ \varepsilon \{\chi , \varphi \}\, . \label{4.10}
\end{eqnarray}
Here $\varepsilon$ is an infinitesimal quantity. The corresponding
gauge generator $\varepsilon \varphi$ generates the infinitesimal
canonical transformations
\begin{eqnarray}
&&{\boldsymbol{q}} \rightarrow {\boldsymbol{q}}+ \delta
{\boldsymbol{q}}\,, \;\;\; {\boldsymbol{p}} \rightarrow
{\boldsymbol{p}} + \delta {\boldsymbol{p}}\,, \;\;\; \delta
{\boldsymbol{q}} = \{\varepsilon \varphi , {\boldsymbol{q}} \}\,,
\;\;\; {\boldsymbol{p}} = \{\varepsilon \varphi , {\boldsymbol{p}}
\}\,,\nonumber \\
&& \bar{\boldsymbol{q}} \rightarrow \bar{\boldsymbol{q}}+ \delta
\bar{\boldsymbol{q}}\,, \;\;\; \bar{\boldsymbol{p}} \rightarrow
\bar{\boldsymbol{p}} + \delta \bar{\boldsymbol{p}}\,, \;\;\;
\delta \bar{\boldsymbol{q}} = \{\varepsilon \varphi ,
\bar{\boldsymbol{q}} \}\,, \;\;\; \bar{\boldsymbol{p}} =
\{\varepsilon \varphi , \bar{\boldsymbol{p}} \}\, . \label{ct1}
\end{eqnarray}
It follows immediately that the corresponding generating function
is
\begin{eqnarray}
G({\boldsymbol{q}},\bar{\boldsymbol{q}}, {\boldsymbol{P}},
\bar{\boldsymbol{P}}) = {\boldsymbol{q}}{\boldsymbol{P}} +
\bar{\boldsymbol{q}}\bar{\boldsymbol{P}} + \varepsilon \varphi +
o(\varepsilon^2)\, .
\end{eqnarray}
The canonical transformations (\ref{ct1}) result in changing
$\varphi$ and $\phi_i$ by
\begin{eqnarray}
&&\delta \varphi \ = \ A \varphi\, , \label{4.18}\\
&&\delta \phi_i \ = \ \varepsilon \{ \phi_i, \varphi\} \ = \ B_i
\varphi + D_{ij} \ \phi_j \, . \label{4.19}
\end{eqnarray}
Here $A, B_i, C$ and $D_{ij}$ are some phase-space functions of
order $\varepsilon$. Note that in our case the gauge algebra is
Abelian\footnote{If $\mathcal{F}$ is any phase-space function then
$[\delta_{\varepsilon}, \delta_{\eta} ] {\mathcal{F}} =
\delta_{\varepsilon}\delta_{\eta} {\mathcal{F}} - \delta_{\eta}
\delta_{\varepsilon} {\mathcal{F}} = \varepsilon\eta \left\{
{\mathcal{F}}, \{ \varphi, \varphi\} \right\} = 0$.}. As a
consequence of (\ref{4.18}) and (\ref{4.19}) we find
\begin{eqnarray}
\delta[\varphi] \ &\rightarrow& \ \left|1 + \mbox{Tr}(A)
\right|^{-1} \delta[\varphi]\, , \label{4.11}\\
\prod_i \delta[\phi_i] \ &\rightarrow& \ \left|1 + \mbox{Tr}
(D)\right|^{-1}
\prod_i \delta[\phi_i]\, , \label{4.12}\\
\sqrt{\left|\det \|\{ \phi_i, \phi_j  \}\| \right|} \
&\rightarrow& \ \left|1 + \mbox{Tr}(D) \right| \ \sqrt{\left|\det
\|\{ \phi_i, \phi_j \}\| \right|}\, . \label{4.13}
\end{eqnarray}
[here $\mbox{Tr}(A) = \sum_t A(t)$, etc.] In (\ref{4.13}) we have
used the fact that in the path-integral measure are present
$\delta[\varphi]$ and $\delta[\phi_i]$, and so we have dropped on
the RHS's of (\ref{4.11})-(\ref{4.13}) the vanishing terms. The
infinitesimal gauge transformations described hitherto clearly show
that $Z_{\rm CM}$ is dependent on the choice of $\chi$ [the term
with $|1+ \mbox{Tr}(A)|$ does not get canceled]. To ensure the gauge
invariance we need to factor out the ``orbit volume" from the
definition of $Z_{\rm CM}$. This will be achieved by a procedure
that is akin to the Faddeev-Popov-De~Witt trick. We define the
functional
\begin{eqnarray}
\left( \triangle_{\chi}\right)^{-1} \ = \ \int {\mathcal{D}}g \
\delta [\chi^{g}]\, , \label{4.14}
\end{eqnarray}
with $\chi^{g}$ representing the gauge transformed $\chi$. The
superscript $g$ in Eq.(\ref{4.14}) denotes an element of the Abelian
gauge group generated by $\varphi$.
We point out that the functional (\ref{4.14}) is manifestly gauge
invariant since
\begin{eqnarray}
\left( \triangle_{\chi^{g'}}\right)^{-1} \ = \  \int
{\mathcal{D}}g \ \delta[\chi^{g'g}]\ = \ \int {\mathcal{D}}(g'g) \
\delta[\chi^{g'g}] \ = \ \left( \triangle_{\chi}\right)^{-1}\, .
\label{4.15}
\end{eqnarray}
The second identity holds because of the invariance of the group
measure under composition, i.e., ${\mathcal{D}}g =
{\mathcal{D}}(g'g)$. Equations (\ref{4.14}) and (\ref{4.15}) allow
us to write ``$1$" as
\begin{eqnarray}
1 \ = \ \triangle_{\chi} \ \delta[\chi] \int {\mathcal{D}}g \, .
\label{4.16}
\end{eqnarray}
To find an explicit form of $\triangle[\chi]$ we can apply the
infinitesimal gauge transformation (\ref{4.10}). Then
\begin{eqnarray}
\chi^g  \ = \ \chi + \varepsilon \{\chi, \varphi \} + C \varphi \
&\Rightarrow \ & \left(\triangle_{\chi}\right)^{-1} \ = \ \int
{\mathcal{D}}\varepsilon \ \delta[\chi + \varepsilon \{\chi ,
\varphi
\} + C \varphi]\, , \nonumber \\
&\Rightarrow \ & \left. \left(\triangle_{\chi}\right)^{-1}
\right|_{\Gamma^*} \ = \ \left|\det\|\{\chi, \varphi
\}\|\right|^{-1} \, , \label{4.29}
\end{eqnarray}
with the obvious notation $\det \| \{ \chi(t), \varphi(t')\} \| =
\prod_t \{ \chi(t), \varphi(t)\}$. Upon insertion of Eq.(\ref{4.16})
into $Z_{\rm CM}$ we obtain
\begin{eqnarray}
&&Z_{\rm CM} \ = \ \int
{\mathcal{D}}{\boldsymbol{p}}{\mathcal{D}}{\boldsymbol{q}}{\mathcal{D}}
\bar{\boldsymbol{p}}{\mathcal{D}}\bar{\boldsymbol{q}} \ \left|
\det \| \{\chi, \varphi  \}\| \right| \sqrt{\left|\det \|\{
\phi_i, \phi_j \}\| \right|} \; \delta[\chi] \delta[\varphi]
\prod_{i=1}^{2N}
\delta[\phi_i] \nonumber \\
&&\mbox{\hspace{4cm}}\times \ \exp\left[ i\! \int_{t_1}^{t_2} dt \
[{\boldsymbol{p}}\dot{\boldsymbol{q}} +
\bar{\boldsymbol{p}}\dot{\bar{\boldsymbol{q}}} - \bar{H}] +
\int_{t_1}^{t_2} dt \ [{\boldsymbol{q}}{\boldsymbol{J}} +
\bar{\boldsymbol{q}}\bar{\boldsymbol{J}}] \right]\, , \label{4.17}
\end{eqnarray}
where the group volume $G_V = \int {\mathcal{D}}g$ has been factored
out as desired. The partition function (\ref{4.17}) is now clearly
(locally) independent of the choice of the gauge constraints $\chi$.
This is because under the transformation (\ref{4.18}) we have
\begin{eqnarray}
 \det \|\{\chi, \varphi \}\|  \ \rightarrow \ \left(1 +
\mbox{Tr}(A) \right)  \det \|\{ \chi + \delta \chi, \varphi \}\|
\, , \label{4.20}
\end{eqnarray}
and hence the partition function $Z_{\rm CM}$ as obtained by
(\ref{4.17}) takes the same form as the untransformed one, but with
$\chi$ replaced by $\chi + \delta \chi$. Because we deal with
canonical transformations it is implicit in our derivation that the
action in the new variables is identical, to within a boundary term,
with the original action. In path integrals this might be
invalidated by the path roughness and related ordering
problems\footnote{In the literature this phenomenon frequently goes
under the name of the Edwards-Gulyaev effect~\cite{EG}.}. For
simplicity's sake we shall further assume that the latter are absent
or harmless. This happens, for instance, when canonical
transformations are linear. In such cases an infinitesimal change in
$\chi$ does not alter the physical content of the theory present in
$Z_{\rm CM}$. This conclusion may generally not be true globally
throughout phase space. Global gauge invariance, however, is
mandatory in our case since we need a global equivalence between the
partition functions $Z_{\rm CM}$ and $Z_{\rm QM}$ and not mere
perturbative correspondence. Thus the potentiality of Gribov's
copies must be checked in every individual problem separately.

\vspace{3mm}

In passing we may notice that if we arrange the constraints in one
set $\{\eta_a \} = \{\chi, \varphi, \phi_i \}$ we can write
(\ref{4.17}) as
\begin{eqnarray}
&&Z_{\rm CM} \ = \ \int
{\mathcal{D}}{\boldsymbol{p}}{\mathcal{D}}{\boldsymbol{q}}{\mathcal{D}}
\bar{\boldsymbol{p}}{\mathcal{D}}\bar{\boldsymbol{q}} \
\sqrt{\left|\det \| \{ \eta_a, \eta_b \} \| \right|} \;
\prod_{a=1}^{2N+2}
\delta[\eta_a] \nonumber \\
&&\mbox{\hspace{4cm}}\times \ \exp\left[ i\! \int_{t_1}^{t_2} dt \
[{\boldsymbol{p}}\dot{\boldsymbol{q}} +
\bar{\boldsymbol{p}}\dot{\bar{\boldsymbol{q}}} - H] +
\int_{t_1}^{t_2} dt \ [{\boldsymbol{q}}{\boldsymbol{J}} +
\bar{\boldsymbol{q}}\bar{\boldsymbol{J}}] \right]\, . \label{4.21}
\end{eqnarray}
By comparison with (\ref{frad}) we retrieve a well known
result~\cite{Sunder,GT}, namely, that the set $\{ \eta_a \}$ of
$2N+2$ constraints can be viewed as a set of second-class
constraints. Thus, by fixing a gauge we have effectively converted
the original system of $2N$ second-class and {\em one} first-class
constraints into $2N+2$ second-class constraints.

\vspace{3mm}

In view of (\ref{2.10}) and (\ref{4.27}), we can perform a
canonical transformation in the full phase space in such a way
that the new variables are: $P_1 = \chi$, $Q_{1+i} = \phi_{2i}$,
$P_{1+i} = \phi_{2i-1}$; $i =1, \ldots, N$. After a trivial
integration over $P_a$ and $Q_{1+i}$ we find that
\begin{eqnarray}
Z_{\rm CM} \ = \ \int
{\mathcal{D}}\bar{\boldsymbol{P}}{\mathcal{D}}\bar{\boldsymbol{Q}}{\mathcal{D}}{Q}_1
\ \left(\delta[\varphi] \left|\det \left|\!\left|\frac{\delta
\varphi}{\delta Q_1}\right|\!\right| \right|\right) \ \exp\left[i
\!\int_{t_1}^{t_2} dt \ \left[ \bar{\boldsymbol{P}}
\dot{\bar{\boldsymbol{Q}}} - K\right]
 + \int_{t_1}^{t_2} dt\ \bar{\boldsymbol{Q}} {\boldsymbol{j}} \right]\, ,
\end{eqnarray}
where $\bar{P}_a$ and $\bar{Q}_a$ are the remaining canonical
variables spanning the $(2N-2)$-dimensional phase space.
To within a time derivative term the new Hamiltonian is done by the
prescription $K(\bar{\boldsymbol{P}}, \bar{\boldsymbol{Q}}, Q_1) =
H(\bar{\boldsymbol{P}}, \bar{\boldsymbol{Q}}, P_1 = 0, Q_1, Q_{1+i}
=0, P_{1+i} =0)$. The sources ${\boldsymbol{j}} $ are
correspondingly transformed sources ${\boldsymbol{J}}$ and
$\bar{{\boldsymbol{J}} }$. Utilizing the identity
\begin{eqnarray}
\delta[\varphi] \left|\det \left|\!\left|\frac{\delta
\varphi}{\delta Q_1}\right| \! \right| \right| \ = \ \delta[Q_1 -
Q_1^*(\bar{\boldsymbol{P}}, \bar{\boldsymbol{Q}}) ]\, ,
\label{4.23}
\end{eqnarray}
we can finally write
\begin{eqnarray}
Z_{\rm CM} \ = \ \int
{\mathcal{D}}\bar{\boldsymbol{P}}{\mathcal{D}}\bar{\boldsymbol{Q}}
\ \exp\left[i \!\int_{t_1}^{t_2} dt \ \left[ \bar{\boldsymbol{P}}
\dot{\bar{\boldsymbol{Q}}} - K^*\right]
 + \int_{t_1}^{t_2} dt\ \bar{\boldsymbol{Q}} {\boldsymbol{j}} \right]\, .
\label{4.22}
\end{eqnarray}
Here $K^*(\bar{\boldsymbol{P}}, \bar{\boldsymbol{Q}}) =
K(\bar{\boldsymbol{P}}, \bar{\boldsymbol{Q}}, Q_1 =
Q_1^*(\bar{\boldsymbol{P}},\bar{\boldsymbol{Q}}))$. In view of
(\ref{D3}) we can alternatively write $Z_{CM}$ as
\begin{eqnarray}
Z_{\rm CM} \ = \ \int
{\mathcal{D}}\bar{\boldsymbol{P}}{\mathcal{D}}\bar{\boldsymbol{Q}}
\ \exp\left[i \!\int_{t_1}^{t_2} dt \ \left[ \bar{\boldsymbol{P}}
\dot{\bar{\boldsymbol{Q}}} - H_+^*\right]
 + \int_{t_1}^{t_2} dt\ \bar{\boldsymbol{Q}} {\boldsymbol{j}} \right]\, ,
\label{4.30}
\end{eqnarray}
where $H_+^* = H_+(\bar{\boldsymbol{P}}, \bar{\boldsymbol{Q}}, Q_1
= Q_1^*(\bar{\boldsymbol{P}},\bar{\boldsymbol{Q}}), P_a = 0,
Q_{1+i} = 0)$. In passing we may notice that $\bar{P}_a$ and
$\bar{Q}_a$ are true canonical variables on the submanifold
$\Gamma^*$ of the initial conditions for Eq.(\ref{4.24}). Indeed,
in terms of a non-canonical system of variables $\{\zeta_i \} =
\{\varphi; \chi; \phi_i; \bar{\boldsymbol{Q}};
\bar{\boldsymbol{P}}\}$ the Poisson bracket of any two {\em
observable} quantities (say $f$ and $g$) on the constraint
manifold $\mathcal{M}$ is
\begin{eqnarray}
\left. \{f, g \} \right|_{\mathcal{M}} \ =  \ \left. \left[
\sum_{a,b}\ \{\zeta_a, \zeta_b \} \ \frac{\partial f}{ \partial
\zeta_a}\frac{\partial g}{ \
\partial \zeta_b}\right] \right|_{\mathcal{M}} \ = \ \sum_{i,j} \ \{ \bar{P}_i, \bar{Q}_j \}
\ \frac{\partial f^*}{\partial \bar{P}_i} \frac{\partial g^*}{
\partial \bar{Q}_j} \ = \ \sum_{i,j} \ \Omega_{ij} \
\frac{\partial f^*}{\partial \bar{\mathcal{Q}}_i}
\frac{\partial g^*}{
\partial \bar{\mathcal{Q}}_j}\, ,
\label{4.91}
\end{eqnarray}
with $ \{ \bar{\mathcal{Q}}_j \} \ = \ \{ \bar{\boldsymbol{Q}};
\bar{\boldsymbol{P}}  \}$ and with
\begin{eqnarray*}
f^*(\bar{\boldsymbol{Q}}, \bar{\boldsymbol{P}}) \ &=& \ f(\varphi
= 0, \chi = 0, \phi_i = 0, \bar{\boldsymbol{Q}},
\bar{\boldsymbol{P}})\, , \nonumber \\
g^*(\bar{\boldsymbol{Q}}, \bar{\boldsymbol{P}}) \ &=& \ g(\varphi
= 0, \chi = 0, \phi_i = 0, \bar{\boldsymbol{Q}},
\bar{\boldsymbol{P}})\, ,
\end{eqnarray*}
representing the physical quantities on ${\mathcal{M}}$. The latter
depend only on the canonical variables $\bar{\boldsymbol{Q}}$ and
$\bar{\boldsymbol{P}}$ which are the independent variables on
$\Gamma^*$. In deriving (\ref{4.91}) we have used the fact that
various terms are vanishing on account of Eqs.(\ref{4.24}) and
(\ref{4.27}). So, for instance, $[\{\varphi, \zeta_i\} \
\partial f/\partial \zeta_i ]|_{\mathcal{M}} = 0$, $\{\varphi_i, \bar{P}_j \} =
0$,  $\{\varphi_i, \bar{Q}_j \} = 0$, $[\{\chi, \zeta_i\} \
\partial f/\partial \chi ]|_{\mathcal{M}} = 0$, etc. The matrix $\Omega_{ij}$
stands for the $(2N-2)\times(2N -2)$ symplectic matrix.

\vspace{3mm}

$Z_{\rm CM}$ as defined by (\ref{4.22})-(\ref{4.30}) does not
generally represent a (classical) deterministic system. This is
because the constraint $\varphi = 0$ explicitly breaks the BRST
invariance of $Z_{\rm CM}$ which (as illustrated in Section~III) is
key in preserving the classical nature of the partition function.
Indeed, using the relations $\{ \chi, \bar{p}_a \} = \{ \chi, p_a -
\bar{q}_a \} = 0$ we immediately obtain
\begin{eqnarray}
\{ \chi, \varphi \} \ = \ \sum_a \left\{ \frac{\partial
\chi}{\partial q_a} \left( \frac{\partial \varphi}{\partial p_a} +
\frac{\partial\varphi}{\partial\bar{q}_a} \right) - \frac{\partial
\chi}{\partial p_a} \frac{\partial \varphi}{\partial q_a}
\right\}\, ,
\end{eqnarray}
which implies that
\begin{eqnarray}
\left. \{ \chi, \varphi \} \right|_{{\mathcal{M}}, \bar{q}_a  =
\lambda_a} \ = \ \sum_a \left\{ \frac{\partial \chi^*}{\partial
q_a} \frac{\partial \varphi^*}{\partial \lambda_a} -
\frac{\partial \chi^*}{\partial \lambda_a} \frac{\partial
\varphi^*}{\partial q_a}\right\} \ \equiv \ \{\chi^*, \varphi^*
\}\, .
\end{eqnarray}
Here the notations $\chi^*({\boldsymbol{q}}, {\boldsymbol{\lambda}})
= \chi(\boldsymbol{q},\boldsymbol{p} =\boldsymbol{\lambda},
\bar{\boldsymbol{q}} = \boldsymbol{\lambda}, \bar{\boldsymbol{p}} =
0)$ and $\varphi^*({\boldsymbol{q}}, {\boldsymbol{\lambda}}) =
\varphi(\boldsymbol{q},\boldsymbol{\lambda}, \boldsymbol{\lambda},
0)$ were used. We also took advantage of the fact that
$\bar{\boldsymbol{q}} = {\boldsymbol{\lambda}}$ as indicated in
Section III. So the generating functional (\ref{4.22}) (or
(\ref{4.30})) can be rewritten as
\begin{eqnarray}
Z_{\rm CM}[{\boldsymbol{J}} = 0] \ =  \ \int
{\mathcal{D}}{\boldsymbol{q}}{\mathcal{D}}{\boldsymbol{\lambda}}
{{\mathcal{D}}\bar{\boldsymbol{c}}} {\mathcal{D}}{\boldsymbol{c}} \
\exp\left[ i \mathcal{S}     \right] \ \delta[\varphi^*]
\delta[\chi^*] \left| \det \| \{\chi^*, \varphi^*  \}\| \right|\, ,
\label{iv72}
\end{eqnarray}
where the integration over the ghost fields was reintroduced for
convenience. By reformulating $Z_{\rm CM}$ in terms of
${\boldsymbol{q}}, {\boldsymbol{\lambda}}, {\boldsymbol{c}}$ and
$\bar{\boldsymbol{c}}$ we can now easily check the BRST invariance.
The BRST transformations (\ref{4.6}) imply that
\begin{eqnarray}
&&\delta_{\rm BRST} \ \varphi^* \ = \ \frac{\partial
\varphi^*}{\partial q_i} \ \bar{\varepsilon} c_i \ = \ -
\bar{\varepsilon} \pounds_{{X_{\mathcal{Q}}}_{\rm BRST}} \
\varphi^*\, , \nonumber \\ &&\bar{\delta}_{\rm BRST} \ \varphi^* \
= \ - \frac{\partial \varphi^*}{\partial q_i} \ \varepsilon
\bar{c}_i \ = \ - \bar{\varepsilon}
\pounds_{{X_{\overline{\mathcal{Q}}}}_{\rm BRST}} \ \varphi^*\, .
\end{eqnarray}
Here ${\pounds_{X_{\mathcal{Q}}}}_{\rm BRST}$ and
${\pounds_{X_{\overline{\mathcal{Q}}}}}_{\rm BRST}$ represent the
Lie derivatives with respect to flows generated by the BRST and
anti-BRST charges, respectively. Analogous relations hold also for
$\chi^*$. Correspondingly, to the lowest order in
$\bar{\varepsilon}$ we can write
\begin{eqnarray}
\delta[\chi^*] \ &\rightarrow& \ | 1 -
\mbox{Tr}(\bar{\varepsilon}{\pounds_{X_{\mathcal{Q}}}}_{\rm BRST})
|^{-1} \ \delta[\chi^*]\, , \nonumber \\
\left| \det \| \{\chi^*, \varphi^*  \}\| \right| \ &\rightarrow& \
| 1 - \mbox{Tr}(\bar{\varepsilon}{\pounds_{X_{\mathcal{Q}}}}_{\rm
BRST}) | \left| \det \| \{\chi^*, \varphi^*  \}\| \right|\, .
\label{iv74}
\end{eqnarray}
The transformations (\ref{iv74}) show that the term $\delta[\chi^*]
 \left| \det \| \{\chi^*, \varphi^*  \}\| \right| $ in
(\ref{iv72}) is the BRST invariant (as, of course, are both the
integration measure and the effective action ${\mathcal{S}}$).
However, because the variation $\delta_{\rm BRST} \delta[\varphi^*]$
is not compensated in (\ref{iv72}) we have in general,  $\delta_{\rm
BRST} Z_{\rm CM}[{\boldsymbol{J}} = 0] \neq 0$. An analogous result
applies also to the anti-BRST transformation.
\vspace{3mm}

We should note that the condition $\delta_{\rm BRST} Z_{\rm
CM}[{\boldsymbol{J}} = 0] \neq 0$ only indicates that the {\em
classical} path-integral structure is destroyed; it does not,
however, ensure that the ensuing $Z_{\rm CM}$ can be recast into a
form describing a proper quantum-mechanical generating functional.
%
%
%
The straightforward path-integral representation such as
(\ref{4.22}) emerges only after the gauge freedom inherent in
the ``information loss" condition $\varphi$ is properly fixed via
the gauge constraint $\chi$. Let us finally emphasize once more that
the partition function (\ref{4.22}) (resp. (\ref{4.30})) has arisen
as a consequence of the application of the classical Dirac-Bergmann
algorithm for singular systems to the classical path integral of
Gozzi {\em et al.}.

%


\section{Explicit examples}\label{SEc5}
\subsection{Free particle}

Although the preceding construction may seem a bit abstract, its
implementation is quite straightforward. Let us now illustrate
this with two systems. As a warm-up example we start with the
Hamiltonian
\begin{eqnarray}
H =L_3= xp_y - yp_x\, , \label{5.1}
\end{eqnarray}
which is known to represent the angular momentum with values
unbounded from below. Alternatively, (\ref{5.1}) can be regarded as
describing the mathematical pendulum. This is because the
corresponding dynamical equation (\ref{eq.1.1.1}) for
${\boldsymbol{q}}$ is a plane pendulum equation with the pendulum
constant $l/\mbox{{\textsl{g}}} =1$. The Lagrangian (\ref{lag1})
reads
\begin{eqnarray}
\bar{L} = \overlinen{x}\dot{x} + \overlinen{y}\dot{y} +
\overlinen{x}y - \overlinen{y}x\, . \label{lag2}
\end{eqnarray}
It is well-known~\cite{Lutzky} that the system has two
(functionally independent) constants of motion - Casimir
functions. For (\ref{5.1}) they read
\begin{eqnarray}
C_1 \ = \ x^2 + y^2\, , \;\;\; C_2 \ = \ xp_x + yp_y\, .
\end{eqnarray}
The charge $C_1$ corresponds to the conserved radius of the orbit
while $C_2$ is the Noether charge of dilatation invariance of the
Lagrangian (\ref{lag2}) under the transformations $(\overlinen{x},
\overlinen{y}, x, y) \mapsto (e^{-s}\overlinen{x},
e^{-s}\overlinen{y}, e^sx, e^sy)$. As only $C_1$ is
$\boldsymbol{p}$-independent, the functions $F_+$ and $F_-$ of
this system are according to Eq.~(\ref{FCH}) chosen as:
\begin{eqnarray}
F_+ \ = \ \frac{(H + a_1 C_1)^2}{4a_1 C_1}\, , \;\;\; F_- \ = \
\frac{(H - a_1 C_1)^2}{4a_1 C_1}\, .
\end{eqnarray}
Hence $H_- = 0 $ implies that $H_+ \approx a_1(x^2 + y^2)$. Here
$a_1$ is some constant to be specified later. The ensuing
first-class constraint is
\begin{eqnarray}
\varphi \ = \ xp_y - yp_x - a_1 x^2 - a_1 y^2 - \bar{p}_{\bar{x}}
\bar{y} + 2a_1 \bar{p}_{\bar{x}} x + \bar{p}_{\bar{y}} \bar{x} +
2a_1 \bar{p}_{\bar{y}} y  \ \approx \ H - a_1 C_1\, .
\end{eqnarray}
The gauge condition can then be chosen in the form $\chi =
\bar{p}_{\bar{y}} - y$. Indeed, we easily find that
\begin{eqnarray}
&&\{ \chi, \varphi \} \ = \ \bar{p}_{\bar{x}} - x \ \neq \ 0\, ,
\nonumber \\
&&\{ \chi, \phi_i \} \ = \ 0\, , \; \; \; i = 1, \ldots, 4\, .
\label{5.5}
\end{eqnarray}
The advantage of our choice of $\chi$ is that it will not run into
Gribov ambiguities, i.e., the equation $\varphi = 0$ will have
globally unique solution for $Q_1$ on $\Gamma^*$. This should be
contrasted with such choices as, e.g., $\chi = p_x$ or $\chi = p_y$,
which also satisfy the conditions (\ref{5.5}), but lead to two
Gribov copies each.

\vspace{3mm}

With the above choice of $\chi$ we may directly write the
canonical transformations:
\begin{eqnarray}
&&P_1 \ = \ \chi \ = \ \bar{p}_{\bar{y}} - y\, , \;\;\; Q_1 \ = \
p_y\, ,\nonumber \\
&&P_2 \ = \ p_x - \bar{x}\, , \mbox{\hspace{1.2cm}} Q_2 \ = \
\bar{p}_{\bar{x}}\, , \nonumber \\
&&P_3 \ = \ p_y - \bar{y}\, , \mbox{\hspace{1.2cm}} Q_3 \ = \
\bar{p}_{\bar{y}}\, , \nonumber \\
&&\bar{P}~ \ = \ \bar{p}_{\bar{x}} - x\, , \mbox{\hspace{1.2cm}}
\bar{Q}~ \ = \ p_x\, .
\end{eqnarray}
 It might be checked that the transformation Jacobian is indeed $1$.
 In the new canonical variables the Hamiltonian $K$ reads
\begin{eqnarray}
K(\bar{P}, \bar{Q}, Q_1) \ = \ H(\bar{P}, \bar{Q}, P_a = 0, Q_1,
Q_2 = 0, Q_3 = 0 ) \ = \ - \bar{P}Q_1 \, .
\end{eqnarray}
The functional $\delta$-function (\ref{4.23}) has the form
\begin{eqnarray}
\delta[Q_1 - Q_1^*(\bar{P},\bar{Q})] \ = \ \delta[Q_1 + a_1
\bar{P}]\, ,
\end{eqnarray}
and hence $K^*(\bar{P}, \bar{Q}) = H_+^*(\bar{P}, \bar{Q})= a_1
\bar{P}^2$. Let us now set $a_1 = 1/2m\hbar$. After changing
variables $\bar{Q}(t)$ to $\bar{Q}(t)/\hbar$ we obtain not only
the correct ``quantum-mechanical" path-integral measure
\begin{eqnarray}
{\mathcal{D}}\bar{{Q}}{\mathcal{D}}\bar{{P}} \ \approx \ \prod_i
\left(\frac{ d {\bar{{Q}}}(t_i) d {\bar{{P}}}(t_i)}{2\pi
\hbar}\right)\, ,
\end{eqnarray}
but also the prefactor $1/\hbar$ in the exponent. So (\ref{4.30})
reduces to the quantum partition function for a free particle of
mass $m$. As the constant $a_1$ represents the choice of units (or
scale factor) for $C_1$ we see that the quantum scale $\hbar$ is
implemented into the partition function via the choice of the ``loss
of information" constraint.

\subsection{Harmonic oscillator}

The system (\ref{5.1}) can also be used to obtain the quantized
linear harmonic oscillator. This is possible by observing that not
only $C_1 = x^2 + y^2$ is a constant of motion for (\ref{5.1}) but
also $C_1 = x^2 + y^2 + c$ with $c$ being any $\boldsymbol{q}$ and
$\boldsymbol{p}$ independent constant. So in particular we can
choose $c = c(\bar{\boldsymbol{q}})$. The functional dependence of
$c$ on $\bar{\boldsymbol{q}}$ cannot be, however, arbitrary. The
requirement that 't~Hooft's constraint should not generate any new
(i.e., secondary) constraint represents quite severe restriction.
Indeed, in order to satisfy Eq.(\ref{D2}) the following condition
must hold (c.f. Appendix D):
\begin{eqnarray}
\sum_{i=0}^{2N} e_i \{\phi_i, \bar{H} \} \ = \ - \sum_{a,i} a_i \{
C_i, \bar{p}_a \} \{ p_a, \bar{H}\} \ = \  \sum_{i,k,a} a_i
\frac{\partial c_i(\bar{\boldsymbol{q}})}{\partial \bar{q}_a}
\bar{q}_k \frac{\partial f_k(\boldsymbol{q})}{\partial q_a}
\end{eqnarray}
which for the system in question is weakly zero only if
\begin{eqnarray}
\bar{x} \frac{\partial c(\bar{\boldsymbol{q}})}{\partial \bar{y}}
- \bar{y} \frac{\partial c(\bar{\boldsymbol{q}})}{\partial
\bar{x}} \ = \ 0\, .
\end{eqnarray}
The latter equation has the solution (modulo irrelevant additive
constant) $c(\bar{\boldsymbol{q}}) = d^2 ({\bar{x}}^2  +
{\bar{y}}^2)$. Here $d^2$ represents a multiplicative constant.
Hence we have that $C_1$ has the general form
\begin{eqnarray}
C_1 \ = \ x^2 + y^2 + d^2({\bar{x}}^2  + {\bar{y}}^2)\, .
\end{eqnarray}
It will be further convenient to choose $a_1 = -1/2d$. The resulting
first-class constraint then reads
\begin{eqnarray}
\varphi \ &=& \ xp_y - yp_x + \frac{1}{2d} x^2 + \frac{1}{2d} y^2
- \frac{d}{2} {\bar{x}}^2 - \frac{d}{2} {\bar{y}}^2 -
\bar{y}\bar{p}_{\bar{x}} + \bar{x} \bar{p}_{\bar{y}} - \frac{1}{d}
x \bar{p}_{\bar{x}} - \frac{1}{d} y \bar{p}_{\bar{y}} + d
\bar{x}p_x + d \bar{y}p_y \nonumber \\
&\approx& \ H + \frac{1}{2d} \ C_1 \, .
\end{eqnarray}
If we choose the gauge condition to be
%
%
\begin{eqnarray}
\chi \ = \  \bar{p}_{\bar{y}} + d p_x- y\, , \label{5.6}
\end{eqnarray}
it ensures that
\begin{eqnarray}
&&\{ \chi, \varphi \} \ = \ 2 \bar{p}_{\bar{x}} - 2 x - 2 d  p_y \
\neq \ 0\, ,\nonumber \\
&&\{ \chi, \phi_i \} \ = \ 0\, , \;\;\; i = 1, \ldots, 4\, .
\end{eqnarray}
In addition, we shall see that (\ref{5.6}) guarantees the unique
global solution of the equation $\varphi = 0$ for $Q_1$ on
$\Gamma^*$ (hence it avoids the undesired Gribov ambiguity).

\vspace{3mm}

The canonical transformation discussed in Section IV now takes the
form
\begin{eqnarray}
&&P_1 \ = \ \chi \ = \ \bar{p}_{\bar{y}} + dp_x - y\, , \;\;\; Q_1
\ = \
p_y\, ,\nonumber \\
&&P_2 \ = \ p_x - \bar{x}\, , \mbox{\hspace{2.15cm}} Q_2 \ = \
\bar{p}_{\bar{x}}\, , \nonumber \\
&&P_3 \ = \ p_y - \bar{y}\, , \mbox{\hspace{2.15cm}} Q_3 \ = \
\bar{p}_{\bar{y}}\, , \nonumber \\
&&\bar{P}~ \ = \ \bar{p}_{\bar{x}} + dp_y- x\, ,
\mbox{\hspace{1.2cm}} \bar{Q}~ \ = \ p_x\, ,
\end{eqnarray}
and the Hamiltonian $K$ reads
\begin{eqnarray}
K(\bar{P}, \bar{Q}, Q_1) \ = \ -\bar{P}Q_1 + d Q_1^2 - d
\bar{Q}^2\, .
\end{eqnarray}
The functional $\delta$-function (\ref{4.23}) now has the form
\begin{eqnarray}
\delta[Q_1 - Q_1^*(\bar{P}, \bar{Q})] \ = \ \delta[Q_1 -
\frac{1}{2d}\ \bar{P}]\, .
\end{eqnarray}
This finally implies that the Hamiltonian on the physical space
$\Gamma^*$ has the form $K^*(\bar{P}, \bar{Q}) = H_+^*(\bar{P},
\bar{Q}) = -(1/4d) \bar{P}^2 - d \bar{Q}^2$. By choosing $d = -
m\hbar/2$ and transforming $\bar{Q} \mapsto \bar{Q}/\hbar$ in the
path integral (\ref{4.22}) (resp. (\ref{4.30})) we obtain the
quantum partition function for a system described by the
Hamiltonian: $(1/2m)\bar{P}^2  + (m/2)\bar{Q}^2$, i.e., the linear
harmonic oscillator with a unit frequency. This is precisely the
result which in the context of the system (\ref{5.1}) was
originally conjectured by 't~Hooft in Ref.~\cite{tHooft3}. Note
again that the fundamental scale (suggestively denoted as $\hbar$)
was implemented into the theory via the ``loss of information"
condition.

\subsection{Free particle weakly
coupled to Duffing's oscillator}

There is no difficulty, in principle, in carrying over our procedure
to non-linear dynamical systems. As an illustration we will consider
here the R\"{o}ssler system. This is a three-dimensional
continuous-time chaotic system described by the three autonomous
nonlinear equations
\begin{eqnarray}
\frac{dx}{dt} &=& -y - z\, ,\nonumber \\
\frac{dy}{dt} &=& x + Ay\, ,\nonumber \\
\frac{dz}{dt} &=& B + xz - Cz\, , \label{5.3}
\end{eqnarray}
where $A$, $B$, and $C$ are adjustable constants. The associated
't\,Hooft Hamiltonian reads
\begin{eqnarray}
H = -p_x(y + z) + p_y(x + Ay) + p_z(B + xz - Cz)\, , \label{5.4}
\end{eqnarray}
and the Lagrangian (\ref{lag1}) has the form
\begin{eqnarray}
\overlinen L = \overlinen{x}\dot{x} + \overlinen{y}\dot{y} +
\overlinen{z}\dot{z} + \overlinen{x}(y + z)
 - \overlinen{y}(x +Ay)  -
\overlinen{z}(B +xz +Cz)\, .
\end{eqnarray}
\noi The R\"{o}ssler system is considered to be the simplest
possible chaotic attractor with important applications in
far-from-equilibrium chemical kinetics~\cite{Ruelle}. It also
frequently serves as a playground for studying, e.g.,
period-doubling bifurcation cycles or Feigenbaum's universality
theory. For the sake of an explicit analytic solution we will
confine ourselves only to the special case when $A = B = C = 0$.
With such a choice of parameters the R\"{o}ssler system can be
expressed in a scalar form as $\dddot{y}  =  y\dot{y}
+\dot{y}\ddot{y} - \dot{y} $ which ensures its
integrability~\cite{Hied}. The latter implies that in this regime
R\"{o}ssler's system does not posses chaotic attractors.

\vspace{3mm}

To proceed further, we should realize that because $C_i$ are
supposed to be ${\boldsymbol{p}}$-independent their finding is
equivalent to specifying the first integrals of the system
(\ref{5.3}) (i.e., functions that are constant along lines of
$(x,y,z)$ satisfying (\ref{5.3})). In other words, the
differential equations (\ref{5.3}) represent a characteristic
system for the differential equation $\{H, C_i\} = 0$. It is
simple to see that the first integrals of the above R\"{o}ssler
system are $x^2 + y^2 + 2z$ and $z e^{-y}$, hence we can identify
$C_1$ and $C_2$ with
\begin{eqnarray}
C_1 \ = \ (x^2 + y^2 + 2z)^2\,  ,\;\;\;\;\; C_2 \ = \ z^2 e^{-2y}\,
.
\end{eqnarray}
The previous choice provides indeed positive and irreducible
charges. The first class constraint $\varphi$ then reads
\begin{eqnarray}
\varphi \ &=& \ -\ p_x(y + z) \ + \ p_y x  \ + \ p_z xz  -  a_1
(x^2 + y^2 + 2z)^2  -  a_2 z^2 e^{-2y} \nonumber \\
&&\ -\ \bar{p}_{\bar{x}}\left( \bar{y} \ + \ \bar{z}z   \ - \ 4
a_1 x (x^2 + y^2 + 2z)\right) \ + \ \bar{p}_{\bar{y}}\left(\bar{x}
\ + \ 4 a_1 y (x^2 + y^2 + 2z)  -
 2a_2 z^2 e^{-2y} \right)\nonumber \\
&& \ + \ \bar{p}_{\bar{z}} \left( \bar{x} \ - \  \bar{z}x \ + \ 4
a_1 (x^2 +
y^2 + 2z) \ + \ 2a_2 z e^{-2y}\right) \, ,\nonumber \\
&\approx& \ H - a_1 C_1 - a_2 C_2 \, .
\end{eqnarray}
Explicit values of $a_1$ and $a_2$ will be fixed in the footnote
$5$. A little algebra shows that the gauge condition $\chi$ can be
selected, for instance, as
\begin{eqnarray}
\chi \ = \ \bar{p}_{\bar{x}} - y\,  .
\end{eqnarray}
Such a choice satisfies the necessary conditions
\begin{eqnarray}
\{ \chi, \varphi \} \ = \ \bar{p}_{\bar{y}} + \bar{p}_{\bar{z}} +
x \ \neq \ 0\, ,\;\;\;\;\;\;\;\;\;\; \{ \chi, \phi_i \} \ = \ 0,
\;\;\; i \ = \ 1, \ldots, 6\, .
\end{eqnarray}
The above $\chi$ also allows us to perform the following linear
canonical transformation:
\begin{eqnarray}
\begin{array}{ll}
P_1 \ = \ \chi \ = \ \bar{p}_{\bar{x}} \ - \ y\, , &~~~~~~ Q_1 \ = \ p_y \, , \\
P_2 \ = \ p_x \ - \ \bar{x}\, , &~~~~~~ Q_2 \ = \
\bar{p}_{\bar{x}}\, ,  \\
P_3 \ = \ p_y \ - \ \bar{y}\, , &~~~~~~ Q_3 \ = \
\bar{p}_{\bar{y}}\, , \\
P_4 \ = \ p_z \ - \ \bar{z}\, , &~~~~~~ Q_4 \ = \
\bar{p}_{\bar{z}}\, ,  \\
\bar{P}_1 \ = \ (\bar{p}_{\bar{z}}/d \ - \ z/d )/\sqrt{2} \, ,
&~~~~~~ \bar{Q}_1 \ = \ (2dp_z \ - \ \bar{p}_{\bar{x}}/c \ + \
x/c)/\sqrt{2} \,
, \\
\bar{P}_2 \ = \ (2cp_x \ - \ \bar{p}_{\bar{z}}/d \ +  \
z/d)/\sqrt{2}\, , &~~~~~~ \bar{Q}_2 \ = \ (x/c \ - \
\bar{p}_{\bar{x}}/c)/\sqrt{2} \, .
\end{array}
\label{5.7}
\end{eqnarray}
Here $c$ and $d$ represent arbitrary real constants to be specified
later. The transformation (\ref{5.7}) secures the unique global
solution $Q_1$ for $\varphi = 0$ on $\Gamma^*$. To show this it is
sufficient to observe that $\left.\left[H - a_1C_1 - a_2C_2
\right]\right|_{\Gamma^*}$ is linear in $Q_1$. Indeed,
\begin{eqnarray}
\left.\left[H - a_1C_1 - a_2C_2 \right]\right|_{\Gamma^*} \ &=& \
\sqrt{2} c \ Q_1 \bar{Q}_2 - \sqrt{2 }c \ (\bar{Q}_1 -
\bar{Q}_2)\bar{Q}_2 \bar{P}_1  + d/c\ (\bar{P}_1 + \bar{P}_2)
\bar{P}_1\nonumber \\ &-& {\mathcal{A}} \ (\bar{P}_1)^2 -
{\mathcal{B}} \ \bar{P}_1 (\bar{Q}_2)^2 - {\mathcal{C}} \
(\bar{Q}_2)^4\, ,
\end{eqnarray}
with ${\mathcal{A}} = 2d^2(4 a_1 + a_2)$, ${\mathcal{B}} = -
8\sqrt{2}a_1  d c^2$ and ${\mathcal{C}} = 4a_1 c^4$. As a result
\begin{eqnarray}
K^*(\bar{\boldsymbol{P}}, \bar{\boldsymbol{Q}}) =
H^*_+(\bar{\boldsymbol{P}}, \bar{\boldsymbol{Q}}) = {\mathcal{A}} \
(\bar{P}_1)^2 + {\mathcal{B}} \ \bar{P}_1 (\bar{Q}_2)^2 +
{\mathcal{C}} \ (\bar{Q}_2)^4\, .
\end{eqnarray}
Inserting this into (\ref{4.22}) (resp. (\ref{4.30})) and
integrating over $\bar{P}_1$ and $\bar{P}_2$ we obtain the following
chain of identities:
\begin{eqnarray}
Z_{\rm CM} \ &=& \ \int {\mathcal{D}}\bar{\boldsymbol{P}}
{\mathcal{D}}\bar{\boldsymbol{Q}} \ \exp\left\{
i\!\!\int_{t_1}^{t_2} dt \ [\bar{\boldsymbol{P}}
\dot{\bar{\boldsymbol{Q}}} - {\mathcal{A}} \ (\bar{P}_1)^2 -
{\mathcal{B}} \ \bar{P}_1 (\bar{Q}_2)^2 - {\mathcal{C}} \
(\bar{Q}_2)^4 +
\bar{\boldsymbol{Q}}{\boldsymbol{j}}~] \right\} \nonumber \\
&=& \ \int {\mathcal{D}}\bar{Q}_1 {\mathcal{D}}\bar{Q}_2 \ \delta
[ \dot{\bar{Q}}_2  ] \ \exp \left\{ i\!\!\int_{t_1}^{t_2} dt \
\left[\frac{1}{4{\mathcal{A}}}\ (\dot{\bar{Q}}_1 - {\mathcal{B}}\
(\bar{Q}_2)^2)^2 - {\mathcal{C}}(\bar{Q}_2)^4 +
\bar{\boldsymbol{Q}}{\boldsymbol{j}}~\right] \right\}\nonumber \\
&=& \ \lim_{a\rightarrow 0_+} \ \int {\mathcal{D}}\bar{Q}_1
{\mathcal{D}}\bar{Q}_2 \ \exp\left\{i\!\! \int_{t_1}^{t_2} dt \
\left[\frac{1}{4 {\mathcal{A}}} (\dot{\bar{Q}}_1)^2 + \frac{1}{4 a}\
(\dot{\bar{Q}}_2)^2 - \frac{\mathcal{B}}{2{\mathcal{A}}} \
\dot{\bar{Q}}_1(\bar{Q}_2)^2\right]
\right\} \nonumber \\
&&\ \mbox{\hspace{2.5cm}} \times \ \exp\left\{ i\!\!
\int_{t_1}^{t_2} dt  \left[\left(
 \frac{{\mathcal{B}}^2}{4{\mathcal{A}}} -
{\mathcal{C}}\right)(\bar{Q}_2)^4 +
\bar{\boldsymbol{Q}}{\boldsymbol{j}}\right] \right\}\, .
\label{5.10}
\end{eqnarray}
As an explanatory step we should mention that the formal measure in
the second equality of (\ref{5.10}) has the explicit time-sliced
form
\begin{eqnarray}
{\mathcal{D}}\bar{Q}_1 {\mathcal{D}}\bar{Q}_2 \ \approx \ \prod_i
\left( \frac{d\bar{Q}_1(t_i)}{\sqrt{4\pi i \epsilon
{\mathcal{A}}}}\ d \bar{Q}_2(t_i) \right)\, ,
\end{eqnarray}
while in the third equality the shorthand notation
${\mathcal{D}}\bar{Q}_1 {\mathcal{D}}\bar{Q}_2$ stands for
\begin{eqnarray}
{\mathcal{D}}\bar{Q}_1 {\mathcal{D}}\bar{Q}_2 \ \approx \ \prod_i
\left( \frac{d\bar{Q}_1(t_i) }{\sqrt{4\pi i \epsilon {\mathcal{A}}}}
\ \frac{d\bar{Q}_2(t_i)}{\sqrt{4 \pi i a \epsilon}}\right)\, .
\end{eqnarray}
The symbol $\epsilon$ represents the infinitesimal width of the
time slicing. During our derivation we have used the Fresnel
integral
\begin{eqnarray}
\int_{-\infty}^{\infty} dx \ e^{-i a x^2 + i x\xi} \ = \
\sqrt{\frac{\pi}{a}}\ \ e^{i (\xi^2/a - \pi)/4 } \ = \
\sqrt{\frac{\pi}{i a}}\ \ e^{i \xi^2/(4a)}\, , \;\;\;\;\;\;\;\;\;
a > 0\, ,
\end{eqnarray}
and the ensuing representation of the Dirac $\delta$-function:
\begin{eqnarray}
\lim _{a\rightarrow 0_{+}}  \sqrt{\frac{1}{4i\pi a}}\ \
e^{i\xi^2/(4a)} \ = \ \delta(\xi)\, . \label{5.11}
\end{eqnarray}
In the following we perform the scale transformation
$\bar{Q}_2/\sqrt{a} \mapsto \sqrt{2m_2}\ \bar{Q}_2$ and set
${\mathcal{A}} = 1/(2m_1)$, ${\mathcal{B}} = 1/(\sqrt{m_1 m_2})$
and ${\mathcal{C}} = 1/m_2$.~\footnote{This choice is equivalent
to the solution:
\begin{eqnarray*} a_1 = \frac{a_2}{4}\, , \;\; d
=\frac{1}{2\sqrt{2a_2m_1}}\, , \;\; c =
\pm\frac{1}{\sqrt[4]{a_2m_2}}\, .
\end{eqnarray*}
Without loss of generality we can set $d = 1/2$, then:
\begin{eqnarray*}
 a_2 = \frac{1}{2m_1}\, , \;\; a_1 = \frac{1}{8m_1}\, , \;\; c = \pm
 2^{3/4} \sqrt[4]{\frac{m_1}{m_2}}\, .
\end{eqnarray*}
}
The resulting partition function then reads
\begin{eqnarray}
Z_{\rm CM} \ &=& \ \lim_{{\rm{g}}\rightarrow 0_+}\int
{\mathcal{D}}\bar{Q}_1 {\mathcal{D}}\bar{Q}_2 \ \exp\left\{ i\!\!
\int_{t_1}^{t_2} dt \left[\frac{m_1}{2}\ (\dot{\bar{Q}}_1)^2 \ + \
\frac{m_2}{2} \ (\dot{\bar{Q}}_2)^2\right]\right\} \nonumber \\
&& \mbox{\hspace{2cm}}\times \ \exp\left\{ i\!\! \int_{t_1}^{t_2}
dt \left[{\rm{g}} \sqrt{\frac{m_1m_2}{2}}
 \ \dot{\bar{Q}}_1(\bar{Q}_2)^2 \ - \ \frac{m_2 {\rm{g}}^2}{4} \
(\bar{Q}_2)^4 \ + \
\bar{\boldsymbol{Q}}{\boldsymbol{j}}\right]\right\}\, ,
 \label{5.9}
\end{eqnarray}
where we have set ${\rm{g}} = 2\sqrt{2} a$.  The system thus
obtained describes a pure anharmonic (Duffing's) oscillator
($\bar{Q}_2$ oscillator) weakly coupled through the Rayleigh
interaction with a free particle ($\bar{Q}_1$ particle).
Alternatively, when $m_1 = m_2 = m$ we can interpret the Lagrangian
in (\ref{5.9}) as a planar system describing a particle of mass $m$
in a quartic scalar potential $e\Phi(\bar{\boldsymbol{Q}}) =  m
{\rm{g}}^2/4\ (\bar{Q}_2)^4$ and a vector potential
$e{\boldsymbol{A}} = ({\rm g}m \sqrt{1/2} \ (\bar{Q}_2)^2, 0)$
(i.e., in the linear magnetic field $B_3 =
\epsilon_{3ij}\partial_iA_j = -  {\rm g} m \sqrt{2} \ \bar{Q}_2/e$).

\vspace{3mm}

It is preferable to set $m_1 \mapsto m_1 \hbar$ and $m_2 \mapsto
m_2/\hbar$. The latter corresponds to the scale factors $a_2 =
1/(2m_1\hbar)$ and $a_1 = 1/(8m_1\hbar)$. After rescaling
$\bar{{Q}}_1(t) \mapsto \bar{{Q}}_1(t)/\hbar$ the partition function
(\ref{5.9}) boils down to the usual quantum-mechanical partition
function with the path-integral measure
\begin{eqnarray}
{\mathcal{D}}{\bar{\boldsymbol{Q}}} \ \approx \ \prod_i
\left(\frac{d\bar{Q}_1(t_i)}{\sqrt{2\pi i  \epsilon \hbar/ m_1}}\
\frac{d \bar{Q}_2(t_i)}{\sqrt{2\pi i \epsilon \hbar/m_2}} \right)\,
,
\end{eqnarray}
and with $1/\hbar$ in the exponent. Hence, just as found in the
previous two cases, the choice of 't~Hooft's condition ensures that
the Planck constant enters the partition function (\ref{5.9}) in a
correct quantum-mechanical manner. In turn, $\hbar$ enters only via
the scale factors $a_1$ and $a_2$ (the factors $d$ and $c$ are
$\hbar$ independent) and hence it represents a natural scale on
which the ``loss of information" condition operates. In other words,
whenever one would be able to ``measure" or determine from ``first
principles" the ``loss of information" condition one could, in
principle, determine the value of the fundamental quantum scale
$\hbar$.

\vspace{3mm}

As a final note we mention that the 't~Hooft quantization procedure
can be straightforwardly extended to other non-linear systems and
particularly to systems possessing chaotic behavior (e.g., strange
attractors). In general cases this might be, however, hindered by
our inability to find the corresponding first integrals (and hence
$C_i$'s) in the analytic form. It is interesting to notice that
machinery outlined above allows to find the emergent quantistic
system for the configuration-space strange attractors. This is
because in 't~Hooft's ``quantization" one only needs the dynamical
equations in the {\em configuration} space. The latter should be
contrasted with the Hamiltonian (or symplectic) systems where
strange attractors cannot exist in the {\em phase-space} on account
of the Liouville theorem~\cite{Hop}.

\section{Conclusions and Outlook}

In this paper we have attempted to substantiate the recent proposal
of G.'t~Hooft in which quantum theory as viewed as not a complete
final theory, but is in fact an emergent phenomenon arising from a
deeper level of dynamics. The underlying dynamics are taken to be
classical mechanics with singular Lagrangians supplied with an
appropriate information loss condition. With plausible assumptions
about the actual nature of the constraint dynamics, quantum theory
is shown to emerge when the classical Dirac-Bergmann algorithm for
constrained dynamics is applied to the classical path integral of
Gozzi {\em et al.}.

\vspace{3mm}

There are essentially two different tactics for implementing the
classical path integrals in 't~Hooft's quantization scenario. The
first is to apply the configuration-space formulation~\cite{GozziI}.
This is suited to situations when 't~Hooft's systems are phrased
through the Lagrangian description. The alternative approach is to
start with the phase-space version~\cite{GozziII}. The latter
provides a natural framework when the Hamiltonian formulation is of
interest or where the language of symplectic geometry is preferred.
It should be, however, stressed that it is not merely a matter of a
computational convenience which method is actually employed. In
fact, both approaches are mathematically and conceptually very
different (as they are also in conventional quantum
mechanics~\cite{Pain,Sh1}). Besides, the methodology for handling
singular systems is distinct in Lagrangian and Hamiltonian
formulations (c.f. Refs.~\cite{Sunder,GT} and citations therein). In
passing, we should mention that the currently popular
Hamilton-Jacobi~\cite{Gu1} and
Legendre-Ostrogradski\u{i}~\cite{Pons} approaches for a treatment of
constrained systems, though highly convenient in certain cases
(e.g., in higher-order Lagrangian systems), have not found as yet
any particular utility in the present context.

\vspace{3mm}

Throughout this paper we have considered only the
configuration-space formulation of classical path integrals.
(Incidently, the phase-space path integral which appears in Section
IV (after Eq.(\ref{3.40})) is not the phase-space path integral {\em
\`{a} la} Gozzi, Reuter and Thacker~\cite{GozziII} but rather
Gozzi's configuration-path~\cite{GozziI} integral with extra degrees
of freedom.) By choosing to work within such a framework we have
been able to render a number of formal steps more tractable (e.g.,
BRST analysis is reputed to be simpler in the configuration space,
uniqueness proof for 't~Hooft systems is easy and transparent in the
Lagrange description, etc.). The key advantage, however, lies in two
observations. First, the position-space path integral of Gozzi {\em
et al.} provides a conceptually clean starting point in view of the
fact that it represents the classical limit of both the
stochastic-quantization path integral and the closed-time-path
integral for the transition probability of systems coupled to a heat
bath. Such a connection is by no means obvious in the canonical
path-integral representation as both the Parisi-Wu stochastic
quantization and the Feynman-Vernon formalism (with ensuing
closed-time-path integral) are intrinsically formulated in the
configuration space. Second, according to 't~Hooft's conjecture the
``loss of information" condition should operate in the position
space where it is supposed to eliminate some of the transient
trajectories leaving behind only stable (or near to stable)
orbits~\cite{tHooft3}. Hence working in configuration space may
allow one to probe the plausibility of 't~Hooft's conjecture. The
price that has been paid for this choice is that the configuration
space must have been doubled. This is an unavoidable step whenever
one wishes to obtain first-order autonomous dynamical equations
directly from the Lagrange formulation (a fact well known in the
theory of dissipative systems~\cite{MF1}). Our analysis in Appendix
BII
suggests, that the auxiliary coordinates $\bar{q}_i$ may be related
to relative coordinates on the backward-forward time path in the
Feynman-Vernon approach. (Such coordinates also go under the names
{\em fast variables}~\cite{Fetter} or {\em quantum noise
variables}~\cite{Sriv1}.)
On the formal side, the auxiliary variables $\bar{q}_i$ are nothing
but Gozzi's Lagrange multipliers ${\lambda}_i$ (in our case denoted
as $\bar{\lambda}_i$).

\vspace{3mm}

In order to incorporate the ``loss of information" into our scheme,
we have introduced in Section IV an auxiliary momentum integration
to go over to the canonical representation. Such a step, though
formal, allowed us to treat our constrained system via the standard
Dirac-Bergmann procedure. It should be admitted that such a choice
is by no means unique - e.g.,  methodologies for treatment of
classical constrained systems in configuration space do
exist~\cite{Sunder,GT}. The decision to apply the Dirac-Bergmann
algorithm was mainly motivated by its conceptual simplicity and
direct applicability to path integrals. On the other hand, we do not
expect that the presented results should undergo any substantial
changes when some another scheme would be utilized. It should be
further emphasized that while we have established the mathematical
link (Eqs.(\ref{4.26}) and (\ref{D3})) between the ``loss of
information" condition and first-class constraints, it is not yet
clear if this connection has more direct physical interpretation
(although various proposals exist in the
literature~\cite{BJV3,tHooft3,BMM1}). Such an understanding would
not only help to develop this approach for more complicated physical
situations but also affiliation in a systematic fashion of a quantum
system to an underlying classical dynamics. Work along those lines
is currently in progress.

\vspace{3mm}

To illustrate the presented ideas we have considered two simple
systems; the planar pendulum and the R\"{o}ssler system. In the
pendulum case we have taken advantage of free choice of an additive
constant in the charge $C_1$. This in turn, allowed us to imposed
't~Hooft's constraints in two distinct ways. In the case of
R\"{o}ssler's system two ${}\boldsymbol{p}$-independent, irreducible
charges $C_1$ and $C_2$ exist. For definiteness sake we have
constructed in the latter case the ``loss of information" condition
with the additive constant set to zero. With this we were able to
convert the corresponding classical path integrals into path
integrals describing a quantized free particle, a harmonic
oscillator, and a free particle weakly coupled to Duffing's
oscillator. As a byproduct we could observe that our prescription
provides a surprisingly rigid structure with rather tight
maneuvering space for the emergent quantum dynamics. Indeed, when
the classical dynamics is fixed, the 't~Hooft condition is
formulated via linear combination of charges $C_i$ which correspond
to the first integrals of the autonomous dynamical equations for
${\boldsymbol{q}}$, i.e., Eq.(\ref{eq.1.1.1}). Due to the explicit
form of 't~Hooft's Hamiltonian the constraint is of the first class
and so we must remove the redundancy in the description by imposing
the gauge condition $\chi$. By requiring that the consistency
conditions (\ref{4.25}) and (\ref{4.27}) are fulfilled, that the
choice of $\chi$ does not induce Gribov ambiguity, and that the
canonical transformations defined in Sec.~IV are linear, we
substantially narrowed down the class of possible emergent quantum
systems. Note also, that when we start with the $N$-dimensional
classical system (${\boldsymbol{q}}$ variables), the emergent
quantum dynamics has $N-1$ dimensions ($\bar{\boldsymbol{Q}}$
variables). Indeed, by introducing the auxiliary degrees of freedom
$\bar{\boldsymbol{q}}$ we obtain $4N$-dimensional phase space which
is constrained by $2N + 2$ conditions ($\phi_i$, $\varphi$ and
$\chi$), which leaves behind $(2N-2)$-dimensional phase space
$\bar{\boldsymbol{Q}}, \bar{\boldsymbol{P}}$. This disparity between
the dimensionality of the classical and emergent quantum systems
vindicates in part the terminology ``information loss" used
throughout the text.

\vspace{3mm}

An important conclusion of this work is that 't~Hooft's quantization
proposal seems to provide a tenable scenario which allows for
deriving certain quantum systems from classical physics.
It should be stressed that although we assumed throughout that the
deeper level dynamics is the classical (Lagrangian or Hamiltonian)
one, there is in principle no fundamental reason that would preclude
starting with more exotic premises. In particular, our conceptual
reasoning would go unchanged if we had begun with Lagrangians
operating over coordinate superspaces (pseudoclassical
mechanics~\cite{BeMa1}) or with the currently much discussed
discrete classical mechanics (i.e., having foam-, fractal-, or
crystal-like configuration space)~\cite{Kleinert?}, etc.~. The only
prerequisite for such approaches is the possibility of formulating a
corresponding variant of Gozzi's path integral, and a method for
implementing the ``loss of information" constraint in such
integrals.

\vspace{3mm}

There are many interesting applications of the above method.
Applications to chaotic dynamical systems especially seem quite
pertinent. After all, central to our reasoning is a (doubled) set of
real first-order dynamical equations\footnote{Non-trivial are only
the equations over actual configuration space. The dynamical
equations for the auxiliary variables $\bar{q}_i$ are linear and
hence they are not relevant in this connection.} which, under
favorable conditions, may by associated with a chaotic dynamics in
the configuration space. We should emphasize that the reader should
not confuse the above with the extensively studied but unrelated
notion of chaos in Hamiltonian systems - we do not deal here with
dynamical equations on symplectic manifolds. This is important, as
Hamiltonian systems forbid {\em per s\`{e}} the existence of
attractive orbits which are otherwise key in 't~Hooft's proposal. In
this respect our approach is parallel with some more conventional
approaches. Indeed, a direct ``quantization" of the equations of
motion -- originally proposed by Feynman~\cite{Dyson} -- is one of
the techniques for tackling quantization of dissipative
systems~\cite{Tar1,HS1}. In field theories this line of reasoning
was recently progressed by Bir\'{o}, M\"{u}ller, and
Matinyan~\cite{BMM1} who demonstrated that quantum gauge field
theories can emerge in the infrared limit of a higher-dimensional
classical (non-Abelian) gauge field theory, known to have chaotic
behavior~\cite{BMM2}.

\vspace{3mm}

We finally wish to comment on two more points. First, in cases where
one strives for an explicit reparametrization invariance (or general
covariance) of the emergent quantum system
the presented framework is not very suitable. The absence of
explicit covariance in both Dirac-Bergmann and Fadeev-Senjanovic
algorithms makes the actual analysis very cumbersome or even
impossible. In fact, expressions (\ref{4.17}) and (\ref{4.21}) are
evidently not generally covariant due to the presence of
time-independent constraints in the measure. Although
generalizations that include covariant constraints do
exist~\cite{fradkin,Bat1,Gav} they result in gauge fixing conditions
which depend not only on the canonical variables but also on the
Lagrange multipliers (or explicit time). Such gauge constraints are,
however, incompatible with our Poisson bracket analysis used in
Section IV, and Appendixes A and D. Hence, if the emergent quantum
system is supposed to be reparametrization invariant (e.g.,
relativistic particle, canonical gravity, relativistic string, etc.)
a new framework for the path-integral implementation of 't~Hooft's
scheme must be sought. Second, the formalism of functional integrals
is sometimes deceptive when taken too literally. The latter is the
case, for instance, when gauge conditions are imposed and/or
canonical transformations performed. The difficulty involved is
known as the Edwards-Gulyaev effect~\cite{Pain,EG,Sh1} and it
resides in the exact nature of the limiting sequence of the finite
dimensional integrals which constitute the path integral. As a
result the classical canonical transformation does not leave, in
general, the measure of the path integral Liouville invariant but,
instead induces an anomaly~\cite{Sh1,SW}. Thus, for our construction
to be meaningful it should be shown that the canonical
transformations in Section IV are unaffected by the Edwards-Gulyaev
effect. Fortunately, in cases when the generating function is at
most quadratic (making canonical transformations  linear) and not
explicitly time dependent, it can be shown~\cite{Fad,SW,vH} that the
anomaly is absent. It was precisely for this reason that more
general transformations were not considered in the present paper.
Clearly, both mentioned points are of key importance for further
development of our procedure and, due to their delicate nature, they
deserve a separate discussion.

\vspace{3mm}

Let us end with the remark that the notorious problem with operator
ordering known from canonical approaches has an elegant solution in
path integrals. The ordering is there naturally generated by the
necessary physical requirement that path integrals must be invariant
under coordinate transformations~\cite{KC}.

\section*{Acknowledgments}

M.B. and P.J. are grateful to the  ESF network COSLAB for funding
their stay at FU, Berlin. One of us, P.J., acknowledges very
helpful discussions with R.~Banerjee,
G.~Vitiello and
Y.~Satoh, and thanks the Japanese Society for Promotion of Science
for financial support.

\section*{Appendix A}

In this appendix we show that the system (\ref{eq.1.1}) has no
secondary constraints. In contract to the primary constraints
which are a consequence of the non-invertibility of the velocities
in terms of the $p$'s and $q$'s, secondary constraints result from
the equations of motion. To show their absence in 't\,Hooft's
system we start with the observation that the time derivative of
any function $f({\boldsymbol{q}}, {\boldsymbol{p}})$ is given
by~\cite{Sunder}
\begin{eqnarray}
\dot{f} \ \approx \ \{ f, \overlinen H \} \  +  \ u^j \{ f,
\phi_j\}\, .
\end{eqnarray}
Here $u^a$ are the Lagrange multipliers to be determined by the
consistency conditions
\begin{eqnarray}
0 \ \approx \ \dot{\phi_i} \ \approx \ \{ \phi_i, \overlinen H \}
\ + \ u^j \{\phi_i, \phi_j \}\, . \label{A1}
\end{eqnarray}
The latter is nothing but the statement that constraints (as
functions of ${\boldsymbol{q}}$ and ${\boldsymbol{p}}$) must hold
at any time. If all $u^j$ could not be determined from the
consistency condition (\ref{A1}) then we would have
 the so-called secondary constraints.  In our case we have
\begin{eqnarray}
\{ \phi_1^a, {\overlinen H} \} \ = \ -  \frac{\partial
\bar{H}}{\partial q_a} \ \not\approx \ 0\, , \;\;\;\; \{ \phi_2^a,
{\overlinen H} \} \ = \ - f_a({\boldsymbol{q}})\ \not\approx \ 0\,
, \;\;\;\; \{ \phi_1^a, \phi_2^b \} \ = \ - \delta_{ab}\, .
\label{A3}
\end{eqnarray}
Using the fact that $\{ \phi_i, {\overlinen H} \} \ \not\approx \
0$ and $\det\left|\{\phi_i, \phi_j \} \right| = 1$, the
inhomogeneous system of linear equations (\ref{A1}) can be
uniquely resolved with respect to $u^j$, thus implying the absence
of secondary constraints.

\section*{Appendix B}
\subsection*{BI}

We show here that Gozzi's configuration-space path integral results
from the ``classical" limit of the stochastic-quantization partition
function, i.e., the limit where the width of a noise distribution
tends to zero.
For this purpose we start with the form of the partition function
for stochastic quantization as written down by
Zinn-Justin~\cite{Zinn-JustinII,Zinn-Justin1}:
 \begin{eqnarray}
Z_{\rm SC}(J) = \int {\mathcal{D}}{\boldsymbol{q}}
{\mathcal{D}}{\boldsymbol{c}} {\mathcal{D}}\bar{{\boldsymbol{c}}}
{\mathcal{D}}\boldsymbol{\lambda} \ \exp\left\{ -
{\mathcal{S}}[{\boldsymbol{q}},{\boldsymbol{c}},\bar{{\boldsymbol{c}}},
\boldsymbol{\lambda}] + \int
{\boldsymbol{J}}(x){\boldsymbol{q}}(x) dx \right\}\, , \label{C1}
 \end{eqnarray}
where
\begin{eqnarray}
{\mathcal{S}} \equiv  &-& w({\boldsymbol{\lambda}}) + \int
{\boldsymbol{\lambda}}(x)\left(
\frac{\partial{\boldsymbol{q}}(x)}{\partial\tau} \ + \frac{\delta
\calS}{\delta {\boldsymbol{q}}(x)} \right)dx \nonumber \\
&-& \int dx dx' \ \bar{c}_a(x) \left( \frac{\partial}{\partial
\tau} \delta_{ab} \delta(x - x') + \frac{\delta^2 \calS }{\delta
q_a(x) \delta q_b(x')}\right) \ c_b(x')\, ,
\end{eqnarray}
and
\begin{eqnarray}
\exp[w({\boldsymbol{\lambda}})]\equiv \int
{\mathcal{D}}{{\boldsymbol{\nu}}}
\exp\left\{-\sigma({\boldsymbol{\nu}}) + \int dx
{\boldsymbol{\lambda}}(x) {\boldsymbol{\nu}}(x)\right\}\, ,
\end{eqnarray}
with  ${\mathcal{D}}{{\boldsymbol{\nu}}}
\exp(-\sigma({\boldsymbol{\nu}}))$ being the functional measure of
noise. Here $x = (t,\tau)$ and $dx = dtd\tau$ where $\tau$ is the
Parisi-Wu fictitious time. The dynamical equation for
${\boldsymbol{q}}(x)$ is described by the Langevin equation
\begin{eqnarray}
\frac{\partial{\boldsymbol{q}}(x)}{\partial \tau} + \left.
\frac{\delta \calS [{\boldsymbol{q}}]}{ \delta
{\boldsymbol{q}}}\right|_{{\boldsymbol{q}} = {\boldsymbol{q}}(x)}
= {\boldsymbol{\nu}}(x)\, , \label{B3}
\end{eqnarray}
with the initial condition ${\boldsymbol{q}}(t,0) =
{\boldsymbol{q}}(t)$.
For Gaussian noise of variance
$2h$, the noise measure is
\begin{eqnarray}
&&{\mathcal{D}}{{\boldsymbol{\nu}}}
\exp(-\sigma({\boldsymbol{\nu}})) =
\prod_{i,x}\frac{d\nu_i(x)}{2\sqrt{\pi \hbar}} \exp\left( -
\frac{1}{4\hbar} \ \int dx {\boldsymbol{\nu}}^2(x) \right)\,
,
\end{eqnarray}
and
(\ref{C1}) takes the form
\begin{eqnarray}
Z_{\rm SC}(J)\ &=& \  \int {\mathcal{D}}{\boldsymbol{q}}
{\mathcal{D}}{\boldsymbol{\nu}} \ \delta \! \left( \frac{\partial
{\boldsymbol{q}}}{\partial \tau} + \frac{\delta \calS
[{\boldsymbol{q}}]}{\delta {\boldsymbol{q}}} -
{\boldsymbol{\nu}}\right) \det\left|\!\left|
\frac{\partial}{\partial \tau} \delta_{ab} \delta(x - x') +
\frac{\delta^2 \calS }{\delta
q_a(x) \delta q_b(x')} \right|\!\right|\nonumber \\
&&\times \ \exp\left\{ - \sigma({\boldsymbol{\nu}}) +   \int
{\boldsymbol{J}}(x){\boldsymbol{q}}(x) dx \right\}\nonumber \\
&=& \ \int {\mathcal{D}}{\boldsymbol{q}}
{\mathcal{D}}{\boldsymbol{\nu}} \ \delta \!\left[{\boldsymbol{q}} -
{\boldsymbol{q}}^{[{\boldsymbol{\nu}}]}\right] \exp\left\{ -
\sigma({\boldsymbol{\nu}}) +   \int
{\boldsymbol{J}}(x){\boldsymbol{q}}(x) dx \right\}\, .
\end{eqnarray}
where $\delta[f({\boldsymbol{q}})] \equiv \prod_{t,\tau}
\delta(f({\boldsymbol{q}}(t,\tau)))$ and
${\boldsymbol{q}}^{[{\boldsymbol{\nu}}]}(x)$ is a solution of
(\ref{B3}). Using the representation.
\begin{eqnarray}
\delta(x) \ = \ \lim_{\hbar \rightarrow 0_+} \frac{1}{2\sqrt{\pi
\hbar}} \ e^{-x^2/(4\hbar)}\, ,
\end{eqnarray}
we get in the limit of zero distribution width (i.e., $\hbar
\rightarrow 0_+$) that
\begin{eqnarray}
Z_{\rm SC}(J,\hbar)\ \rightarrow \ \int
{\mathcal{D}}{\boldsymbol{q}} \ \delta \!\left[{\boldsymbol{q}} -
{\boldsymbol{q}}^{[0]}\right]
 \exp\left\{ \int
{\boldsymbol{J}}(x){\boldsymbol{q}}(x) dx \right\}\, .
\end{eqnarray}
Choosing a special source ${\boldsymbol{J}}(x) = {\boldsymbol{J}}(t)
\delta(\tau)$ we can sum in the path integral solely over
configurations with ${\boldsymbol{q}}(t,0) = {\boldsymbol{q}}(t)$ as
other configurations will contribute only to an overall
normalization constant. Inasmuch we finally obtain
\begin{eqnarray}
\lim_{\hbar \rightarrow 0^+} Z_{\rm SC}({\boldsymbol{J}},\hbar) \ =
\ {\calZ }_{\rm CM}({\boldsymbol{J}})\, .
\end{eqnarray}

\subsection*{BII}

In this part of the appendix we show that Gozzi's
configuration-space partition function (\ref{4.3}) results from the
``classical" limit of the closed-time path integral for the
transition probability of a system coupled to a thermal reservoir at
some temperature $T$. By the classical limit we mean the high
temperature and weak heat bath coupling limit.

The path-integral treatment of systems that are linearly coupled to
a thermal bath of harmonic oscillators was first considered by
Feynman and Vernon~\cite{FV}. For our purpose it will be
particularly convenient to utilize the so called Ohmic limit
version, as discussed in Refs.\cite{Pain,Klein1}:
\begin{eqnarray}
{\mathcal{Z}}_{\rm FV}[{\boldsymbol{J}}_+, {\boldsymbol{J}}_-] \
&=& \ \int {\mathcal{D}}{\boldsymbol{q}}_+
{\mathcal{D}}{\boldsymbol{q}}_- \ \exp\left\{ \frac{i}{\hbar}
\left[ {\mathcal{A}}[{\boldsymbol{q}}_+] -
{\mathcal{A}}[{\boldsymbol{q}}_-] \right]+ \int dt \ \left[
{\boldsymbol{J}}_+(t){\boldsymbol{q}}_+(t)   -
{\boldsymbol{J}}_-(t){\boldsymbol{q}}_-(t)\right]
\right\}\nonumber
\\
&& \times \ \exp\left\{ -i \frac{m\gamma}{2\hbar} \int dt \
[{\boldsymbol{q}}_+(t) -
{\boldsymbol{q}}_-(t)][\dot{{\boldsymbol{q}}}_+(t) +
\dot{{\boldsymbol{q}}}_-(t)]^R \right\}\nonumber \\
&& \times \ \exp\left\{ - \frac{m\gamma}{\hbar^2 \beta} \int dt
\int dt' \ [{\boldsymbol{q}}_+(t) -
{\boldsymbol{q}}_-(t)]K(t,t')[{\boldsymbol{q}}_+(t') -
{\boldsymbol{q}}_-(t')] \right\}\, . \label{B13}
\end{eqnarray}
Here the paths ${\boldsymbol{q}}_+(t)$ and ${\boldsymbol{q}}_-(t)$
are associated with the forward and backward movement of the
particles in time. The super-script $R$ indicates a {\em negative}
shift in the time argument of the velocities with respect to
positions. The latter ensures the causality of the friction
forces~\cite{Klein1}. In addition, $m$ represents the particle mass
(for simplicity we assume here that all system particles have the
same mass), $\beta = 1/T$, and $\gamma$ is the friction constant (or
thermal reservoir coupling). The function $K(t,t')$ is the bath
correlation function. As argued in~\cite{Pain,Klein1}, at high
temperatures $K(t,t') \approx \delta(t-t')$. Introducing the new set
of variables ${\boldsymbol{q}} = [{\boldsymbol{q}}_+ +
{\boldsymbol{q}}_-]/2$ and $\bar{{\boldsymbol{q}}} =
[{\boldsymbol{q}}_+ - {\boldsymbol{q}}_-]$ (i.e., the center-of-mass
and {\em fast} coordinates) we can in the high-temperature case
recast (\ref{B13}) into
\begin{eqnarray}
{\mathcal{Z}}_{\rm FV}[{\boldsymbol{J}}, \bar{{\boldsymbol{J}}}] \
&=& \ \int {\mathcal{D}}{\boldsymbol{q}} {\mathcal{D}}
\bar{{\boldsymbol{q}}} \ \exp\left\{ \frac{i}{\hbar} \left[
{\mathcal{A}}[{\boldsymbol{q}} + \bar{{\boldsymbol{q}}}/2] -
{\mathcal{A}}[{\boldsymbol{q}} - \bar{{\boldsymbol{q}}}/2] \right]
+ \int dt \ \left[{\boldsymbol{J}}(t){\boldsymbol{q}}(t) -
\bar{{\boldsymbol{J}}}(t)\bar{{\boldsymbol{q}}}(t)
\right] \right\}\nonumber \\
&& \times \ \exp\left\{- i\frac{m\gamma}{\hbar} \int dt \
\bar{{\boldsymbol{q}}}(t)\left[\dot{{\boldsymbol{q}}}(t)\right]^R
- \frac{m\gamma}{\hbar^2 \beta} \int dt \
\bar{{\boldsymbol{q}}}^2(t) \right\}\, .
\end{eqnarray}
Here the self-explanatory notation ${\boldsymbol{J}}=
[{\boldsymbol{J}}_+ - {\boldsymbol{J}}_-]$ and
$\bar{{\boldsymbol{J}}} = -[{\boldsymbol{J}}_+ +
{\boldsymbol{J}}_-]/2$  was used. Let us now define $\omega =
2m\gamma /\beta$, integrate over $\bar{{\boldsymbol{q}}}$,
 and go to the
  classical limit $\gamma \rightarrow 0 $. Then we obtain the
following chain of equations:
\begin{eqnarray}
&&\lim_{\gamma \rightarrow 0} \ {\mathcal{Z}}_{\rm
FV}[{\boldsymbol{J}},
\bar{{\boldsymbol{J}}}]\nonumber \\
&& \mbox{\hspace{1cm}}= \ \lim_{\gamma \rightarrow 0} \int
{\mathcal{D}}{{\boldsymbol{q}}}
{\mathcal{D}}\bar{{\boldsymbol{q}}} \ \exp\left\{ \frac{i}{\hbar}
\int dt \ \bar{{\boldsymbol{q}}}(t) \left[
\frac{\delta{\mathcal{A}}}{\delta {\boldsymbol{q}}(t)} - m\gamma
\left[\dot{{\boldsymbol{q}}}(t)\right]^R + i \hbar
\bar{{\boldsymbol{J}}}(t) \right] - \frac{\omega}{2\hbar^2} \int
dt \ \bar{{\boldsymbol{q}}}^2(t) \right\}
\nonumber \\
&& \mbox{\hspace{1.5cm}}\times \ \exp\left\{ \int dt \
{\boldsymbol{J}}(t){\boldsymbol{q}}(t) \right\}\nonumber \\
&& \mbox{\hspace{1cm}}= \ \lim_{\gamma  \rightarrow 0} \ \int
{\mathcal{D}}{{\boldsymbol{q}}} \ \exp \left\{- \frac{1}{2\omega}
\int dt \ \left[ \frac{\delta{\mathcal{A}}}{\delta
{\boldsymbol{q}}(t)} - m\gamma
\left[\dot{{\boldsymbol{q}}}(t)\right]^R + i \hbar
\bar{{\boldsymbol{J}}}(t) \right]^2 +  \int dt \
{\boldsymbol{J}}(t){\boldsymbol{q}}(t)  \right\} \nonumber \\
&& \mbox{\hspace{1cm}}= \ \lim_{\gamma  \rightarrow 0} \ \int
{\mathcal{D}}{{\boldsymbol{q}}} \ {\mathcal{J}}[{\boldsymbol{q}}]\
\exp \left\{- \frac{1}{2\omega} \int dt \ \left[
\frac{\delta{\mathcal{A}}}{\delta {\boldsymbol{q}}(t)} - m\gamma
\dot{{\boldsymbol{q}}}(t) + i \hbar \bar{{\boldsymbol{J}}}(t)
\right]^2 +  \int dt \
{\boldsymbol{J}}(t){\boldsymbol{q}}(t)  \right\} \nonumber \\
&& \mbox{\hspace{1cm}}= \ \int {\mathcal{D}}{{\boldsymbol{q}}} \
\delta \!\left[ \frac{\delta{\mathcal{A}}}{\delta
{\boldsymbol{q}}} + i \hbar \bar{{\boldsymbol{J}}}  \right]
{\mathcal{J}}[{\boldsymbol{q}}] \ \exp\left\{ \int dt \
{\boldsymbol{J}}(t){\boldsymbol{q}}(t) \right\}
\nonumber \\
&& \mbox{\hspace{1cm}}= \ \int {\mathcal{D}}{{\boldsymbol{q}}} \
\delta \!\left[ {\boldsymbol{q}} -
{\boldsymbol{q}}^{[\bar{{\boldsymbol{J}}}]}\right] \ \exp\left\{
\int dt \ {\boldsymbol{J}}(t){\boldsymbol{q}}(t) \right\} \, .
\end{eqnarray}
The Jacobian ${\mathcal{J}}[{\boldsymbol{q}}]$ results from
transition to the ``unretarded" velocities and its explicit form
reads~\cite{Klein1}:
\begin{eqnarray}
 {\mathcal{J}}[{\boldsymbol{q}}] \ = \ \det \left|\!\left|\frac{\partial}{\partial t} \
\delta_{ab}\delta(t-t') + \frac{\delta^2 {\mathcal{A}}}{\delta
q_a(t) \delta q_b(t')}\right|\!\right|\, .
\end{eqnarray}
Coordinates ${\boldsymbol{q}}^{[\bar{{\boldsymbol{J}}}]}$ are
solutions of the equation of the motion:
\begin{eqnarray}
\frac{\delta{\mathcal{A}}[{\boldsymbol{q}}]}{\delta
{\boldsymbol{q}}(t)} \ = \ - i \hbar \bar{{\boldsymbol{J}}}(t)\, .
\end{eqnarray}
In the limit $ \gamma \rightarrow 0$, we find again the Gozzi {\em
et al.} partition function
\begin{eqnarray}
\lim_{\gamma \rightarrow 0} {\mathcal{Z}}_{\rm
FV}[{\boldsymbol{J}}, {{\boldsymbol{0}}}] \ = \ \lim_{\hbar
\rightarrow 0} \lim_{\gamma \rightarrow 0} {\mathcal{Z}}_{\rm
FV}[{\boldsymbol{J}}, \bar{{\boldsymbol{J}}}] \ = \ Z_{\rm
CM}[{\boldsymbol{J}}]\, .
\end{eqnarray}
%
%
%

\section*{Appendix C}

In this appendix we prove that (\ref{3.33}) is a special case of the
Euler-like functionals (\ref{3.31}). Let us first show that
 (\ref{3.33}) can be replaced by
an action of the form (\ref{3.31}). Indeed, because of the
homogeneity of (\ref{3.33}), we can immediatley replace it by
\begin{eqnarray}
 \calS[r^{\alpha_i}q_i] = \sum_i \int dt \ \alpha_i
r^{\alpha_i}(t)q_i(t) \frac{\delta \calS[r^{\alpha_i}q_i]}{\delta
r^{\alpha_i}(t)q_i(t)} \ = \ \int dt \ r(t) \frac{\delta
\calS[r^{\alpha_i}q_i]}{\delta r(t)}\, . \label{B11}
\end{eqnarray}
%
Since this is true for any $r(t)$, we see that
\begin{eqnarray}
\int dt dt' \  r(t) \frac{\delta^2 \calS[r^{\alpha_i}q_i]}{\delta
r(t) \delta r(t')} \ = \ 0\, . \label{B1}
\end{eqnarray}
This simply expresses the fact that the functional $
\calS[r^{\alpha_i}q_i]$ is linear in $r(t)$. The right-hand side of
(\ref{B11}) has then precisely the Euler form (\ref{3.31}).

\vspace{3mm}

The reverse direction is proved in the following way: We first
recast (\ref{3.31}) in the general form
\begin{eqnarray}
\int dt \ r(t) L({\boldsymbol{q}}(t), \dot{{\boldsymbol{q}}}(t)) \ =
\ \int dt \ L\!\left( r^{\alpha_i}(t)q_i(t),
d(r^{\alpha_i}(t)q_i(t))/dt \right)\, . \label{B2}
\end{eqnarray}
Applying the variation  $\int dt \ \delta/\delta r(t)$ to (\ref{B2})
we obtain
\begin{eqnarray}
\calS[{\boldsymbol{q}}]\ = \ \int dt \ \sum_i \alpha_i r^{\alpha_i
-1} q_i(t) \left( \frac{\partial L}{\partial r^{\alpha_i}(t) q_i(t)}
- \frac{d}{dt}\frac{\partial L}{\partial [d(r^{\alpha_i}(t)
q_i(t))/dt]} \right)\, .
\end{eqnarray}
This relation must hold for all $r(t)$, and hence by choosing $r(t)
= 1$ we arrive at the required result
\begin{eqnarray}
\calS[{\boldsymbol{q}}]\ = \ \int dt \ \sum_i \alpha_i q_i(t)
\frac{\delta \calS[{\boldsymbol{q}}]}{\delta q_i(t)}\, .
\end{eqnarray}

\section*{Appendix D}

Here we prove the fact that inclusion of the subsidiary constraint
(\ref{4.1}) in the primary constraints (\ref{2.5}) does not
produce any secondary constraints. The secondary constraints
result from the consistency conditions (\ref{A1}) or, in other
words, when existent constraints are incompatible with the
equation of motion.

\vspace{3mm}

We first observe that the condition $H_- \approx 0$ can be
equivalently represented by the condition $(\bar{H} - \sum_i a_i
C_i) \equiv \phi_0 \approx 0$. If we now add the subsidiary
constraint $\phi_0$ to the remaining $2N$ constraints $\phi_i$ and
again require that the constraints $\phi_i$ remain (weakly) zero
at all times we have
\begin{eqnarray}
0 \ \approx \ \dot{\phi_i} \ \approx \ \{ \phi_i, {\overlinen H}\}
\ + \ u^j \{\phi_i, \phi_j \}\, , \;\;\;\; \;\;\; i,j = 0, 1
\ldots, 2N\, . \label{D1}
\end{eqnarray}
Since there is an odd number of constraints and because $\{\phi_i,
\phi_j \}$ is an antisymmetric matrix we have that $\det\|\left\{
\phi_i, \phi_j \right\}\| = 0$. From the analysis in Appendix A it
is clear that the rank of the matrix $\{\phi_i, \phi_j \}$ is $2N$
and hence it has one null-eigenvector, say ${\boldsymbol{e}}$.
Inasmuch, Eq.(\ref{D1}) implies the constraint
\begin{eqnarray}
\sum_{i=0}^{2N} e_i \{\phi_i, \bar{H} \} \ \approx \ 0\, .
\label{D2}
\end{eqnarray}
If the latter would represent a new non-trivial constraint (i.e.,
constraint that cannot be written as a linear combination of
constraints $\phi_i$) we would need to include such a new
constraint (the so called secondary constraint) into the list of
existent constraints and go again through the consistency
condition (\ref{D1}). Fortunately, the condition (\ref{D2}) is
automatically fulfilled and hence it does not constitute any new
constraint. Indeed, be choosing
\begin{eqnarray}
{\boldsymbol{e}} \ = \ \left(\begin{array}{c} 1 \\
\{\phi_0, \phi_2^a \}\\
\{\phi_1^a, \phi_0\}\\
\{\phi_0, \phi_2^b\}\\
\{\phi_1^b, \phi_0\}\\
\vdots\\
\{ \phi_0, \phi_2^N\}\\
\{ \phi_1^N, \phi_0\}
\end{array}\right) \ = \ \left( \begin{array}{c}
1 \\
{f}_a({\boldsymbol{q}})\\
- \frac{\partial
\phi_0}{\partial q_a} \\
{f}_b({\boldsymbol{q}})\\
-\frac{\partial
\phi_0}{\partial q_b}\\
\vdots\\
{f}_N({\boldsymbol{q}})\\
-\frac{\partial \phi_0}{\partial{q}_N}
\end{array}\right)\, ,
\end{eqnarray}
and using $\{ \phi_0, \bar{H} \} = 0$ together with (\ref{A3}) we
obtain
\begin{eqnarray}
\sum_{i=0}^{2N} e_i \{\phi_i, \bar{H} \}\ = \ - \sum_{i,a} a_i(t)
{f}_a({\boldsymbol{q}}) \ \frac{\partial C_i}{\partial q_a} \ = \
\sum_{i =1}^n a_i(t) \{H, C_i\}\ = \ 0\, . \label{D4}
\end{eqnarray}
As the latter is zero (even strongly) there is no new constraint
condition generated by an inclusion of $\phi_0$ in the original
set of (primary) constraints. Note, that the key in obtaining
(\ref{D4}) was the fact that $C_i$'s are
${\boldsymbol{p}}$-independent constants of motion.

\vspace{3mm}

The rank of $\{ \phi_i, \phi_j \}$ being $2N$ means that there is
one relation
\begin{eqnarray}
\sum_{i = 0}^{2N} e_i \{ \phi_i, \phi_j \} \ \approx \ 0\, .
\end{eqnarray}
Any linear combination of the constraints $\phi_i$ is again a
constraint. So, particularly if we define $\varphi = \sum_i e_i
\phi_i$ we obtain that $\varphi$ has weakly vanishing Poisson
brackets with all constraints, i.e.,
\begin{eqnarray}
\{ \varphi, \phi_i \} \ \approx \ 0\, , \;\;\;\; i = 1, \ldots,
2N\, .
\end{eqnarray}
Thus, according to Dirac's classification (see e.g.,
Ref.~\cite{Dir}) $\varphi$ is a first class constraint. The
remaining $2N$ constraints (which do not have vanishing Poisson
brackets with all other constraints) are of the second class. Note
particularly that the explicit form for $\varphi$ reads
\begin{eqnarray}
\varphi \ = \ \sum_{i = 0}^{2N} e_i \phi_i \ = \ (H - \sum_{i =
1}^n a_i C_i) - \sum_{a = 1}^N \bar{p}_a \ \frac{\partial
\phi_0}{\partial q_a}\, , \label{D3}
\end{eqnarray}
which is clearly weakly identical to $H- \sum_i a_i C_i$. Observe
that it is $H$ and not $\bar{H}$ that is present in (\ref{D3}).

\section*{References}
%


\begin{thebibliography}{99}
\bibitem{Hied} we wish to thank Prof J.~Heidel and Dr Fu~Zhang for clarifying this
point to us.
\end{thebibliography}
\end{document}